\documentclass[11pt,a4paper]{article}

\usepackage[T2A]{fontenc}
\usepackage[cp1251]{inputenc}
\usepackage[english]{babel}

\usepackage[psamsfonts]{amsfonts}
\usepackage{amsthm}
\usepackage{amssymb}
\usepackage{amsmath}
\usepackage{amsopn}
\usepackage{bm}
\usepackage{cite}

\newcommand{\nameABLemma}{$\mathcal{A}$-$\mathcal{B}$ bracket lemma}
\newcommand{\nameConjVLemma}{Conjugate variable lemma}
\newcommand{\nameSoV}{Separation of variables theorem}

\newtheorem{ABLemma}{\nameABLemma}{\bf}{\it}

\newtheorem{ConjVLemma}{\nameConjVLemma}{\bf}{\it}
\newtheorem{SoV}{\nameSoV}{\bf}{\it}

\theoremstyle{plain}
\newtheorem{proposition}{Proposition}[section]

\newtheorem*{proposition*}{Proposition}

\newtheorem{example}{Example} 
\newtheorem{remark}{Remark}

\newcommand{\tens}[1]{\textsf{#1}}
\renewcommand{\vec}[1]{\bm{#1}}

\DeclareMathOperator{\Complex}{\mathbb{C}}

\DeclareMathOperator{\Integer}{\mathbb{Z}}
\DeclareMathOperator{\Tr}{Tr} \DeclareMathOperator*{\res}{res}
\DeclareMathOperator{\spanOp}{span} \DeclareMathOperator{\ad}{ad}
\DeclareMathOperator{\Sym}{Sym} 
\DeclareMathOperator{\const}{const}
\DeclareMathOperator{\diag}{diag}
\DeclareMathOperator{\Ibb}{\mathbb{I}}

\DeclareMathOperator{\rmi}{i}

\allowdisplaybreaks[4]

\begin{document}

\date{}
\title{\bf Orbit Approach to Separation of Variables \\ in $\mathfrak{sl}(3)$-Related Integrable Systems}

\author{\textbf{Julia Bernatska and \fbox{Petro Holod}} \\ \small
 National University of `Kyiv Mohyla Academy'\\ \small
 BernatskaJM@ukma.kiev.ua
 }

\maketitle
\begin{abstract}
Using the orbit method we attempt to reveal geometric and algebraic meaning
of separation of variables for integrable systems
on coadjoint orbits of an $\mathfrak{sl}(3)$ loop algebra. We consider two types of
generic orbits, embedded into a common manifold endowed with two nonsingular
Lie-Poisson brackets. We prove that separation of variables on orbits of both types
is realized by the same variables of separation.
We also construct integrable systems on the orbits:
a coupled 3-component nonlinear Schr\"{o}dinger equation
and an isotropic SU(3) Landau-Lifshitz equation.
\end{abstract}

\section{Introduction}
Separation of variables is a powerful technique for solving problems in mathematical physics,
in particular in the theory of integrable  Hamiltonian systems.
There are well known separation of variables schemes for the integrable systems of KdV and MKdV equations,
sin(sinh)-Gordon equations, nonlinear Sch\"{o}dinger equation, isotropic Landau-Lifshitz equation,
XXX and XYZ magnetic chains, XXX Gaudin model. Many authors contributed to progress of
this technique, the reader can find the corresponding references in Sklyanin's paper \cite{Sklyanin95}
giving, in particular, a review on separation of variables,
and its applications to some of the mentioned systems.

Among all the papers cited by Sklyanin we find a common effective algorithm
linked to the Lax representation of an integrable system. The algorithm fits well into the orbit method
\cite{Kirillov} used in \cite{Adams,AdamsCO}, and developed in the present paper.
The orbit approach gives an opportunity to understand
geometric and algebraic meaning of this separation of variables scheme.

We mention as well another scheme \cite{Blaszak,Magri00R,Magri00T,Falqui} considering
separation of variables within the bi-Hamiltonian geometry.
The manifold where a system lives is called bi-Ha\-mil\-to\-ni\-an if
endowed with a pair of holomorphic Poisson brackets, at least one is nonsingular \cite{Magri}.
Such manifold possesses a set of Nijenhuis or Darboux-Nijenhuis coordinates,
canonical with respect to the corresponding symplectic form. Bi-Hamiltonian property
gives the separability criterion:  integrals of motion, defined as invariant functions,
are in involution with respect to the both Poisson brackets.
The bi-Hamiltonian property gives also an algorithm
of computing variables of separation and exhibits separation relations.
Many, if not most, of the known integrable systems are bi-Hamiltonian.

We distinguish between these two schemes due to different mathematical apparatus.
The first scheme substantially uses the apparatus of loop Lie algebras, but the second deals
with symplectic manifolds not considering them as subspaces of loop algebras.
A connection between these two schemes is shown in \cite{Harnad}.

In the present paper we develop the first scheme of separation of variables
by means of the orbit method. This method deals with integrable systems constructed
on coadjoint orbits in loop algebras. The orbit approach to separation of variables
on orbits of $\mathfrak{sl}(2)$ loop algebras is presented in \cite{BernHolod07}.
Here we develop this approach for systems
on orbits of an $\mathfrak{sl}(3)$ loop algebra, which we call $\mathfrak{sl}(3)$-related systems.
Some of such systems are considered in the cited papers:
the SL(3) magnetic chain in \cite{Sklyanin92}, the 3-particle Calogero-Moser model in \cite{Sklyanin95}, and a coupled 2-component nonlinear Schr\"{o}dinger equation in \cite{Adams}.

The main idea of Sklyanin's paper \cite{Sklyanin95} is the following:
the poles of the properly normalized Baker-Akhiezer function give variables of separation.
This receipt is good for already solved systems,
when a solution in the form of Baker-Akhiezer function is known. Otherwise, one has some uncertainty
caused by the proper normalization. However Sklyanin proposes a procedure of constructing variables of separation.
This procedure is a brilliant idea, and is developed in \cite{Scott, Gekhtman} for
the $\mathfrak{sl}(n)$ case. At the same time Sklyanin declared in \cite{Sklyanin95}
that `algebraic structures underlying the separation of variables' remains unclear.
Attempting to reveal geometric and algebraic meaning of this separation of variables
procedure we appeal to the orbit method. This approach solves the problem of proper normalization
because the relations producing variables of separation are explicitly constructed from a loop algebra
restricted to an orbit. Evidently, these variables serve as poles of the corresponding
Baker-Akhiezer function.

In \cite{Adams,AdamsCO} Adams, Harnad and Hurtubise also solved the problem of
constructing Darboux coordinates
for integrable Hamiltonian systems on coadjoint orbits in loop algebras. These coordinates
was obtained from a divisor of a section of the dual eigenvector
line bundle over the invariant spectral curve of a system. This approach gives some geometric
explanation of the separation of variables scheme. In the present paper we come to similar results,
but propose another explanation that looks intuitively obvious.

In our paper we deal with bi-Hamiltonian systems.
Both the systems in question evolve in a common symplectic manifold, endowed
with two nonsingular Poisson brackets possessing a common spectral curve, and
all integrals of motion are in involution with respect to both the Poisson brackets. The
integrals of motion divides into Casimir functions and Hamiltonians,
Casimir functions with respect to the first Poisson tensor
serve as Hamiltonians for the second Poisson tensor,
and vice verse \cite{Magri}. Therefore, we have two transversal foliations
of the manifold into orbits, every orbit serves as a symplectic leaf.

The paper is organized as follows. After introducing some algebraic structures in Section\;2 we
consider two types of orbits of the standard graded $\mathfrak{sl}(3)$ loop algebra.
These orbits serve as phase spaces for two integrable systems, as shown in Section 3 where
Lie-Poisson brackets and invariant functions are introduced.
Section~4 is devoted to separation of variables on orbits of the first type.
The obtained variables of separation are proven to be canonically conjugate.
Some words about connection to the well-known results are given in Section 5.
Separation of variables on orbits of the second type is realized in Section~6.
We obtain the same relations producing variables of separation,
but the variables are quasi-canonically conjugate on orbits of the second type.
In Section 7 we construct integrable systems on orbits of the both types.
This is a coupled 3-component nonlinear Schr\"{o}dinger equation for the first type,
and an isotropic SU(3) Landau-Lifshitz equation for the second one.
Some conclusion and discussion are given in Section~8.

\section{Preliminaries}
At the beginning we construct a loop algebra based
on the algebra $\mathfrak{g}\,{=}\,\mathfrak{sl}(3,\Complex)$ with the
Cartan-Weyl basis
\begin{gather*}
  \tens{X}_1=\begin{pmatrix}0&1&0\\0&0&0\\0&0&0\end{pmatrix}, \quad
  \tens{X}_2=\begin{pmatrix}0&0&0\\0&0&1\\0&0&0\end{pmatrix}, \quad
  \tens{X}_3=\begin{pmatrix}0&0&1\\0&0&0\\0&0&0\end{pmatrix}, \quad
  \tens{H}_1=\frac{1}{3}\diag(2,-1,-1), \\
  \tens{Y}_1=\begin{pmatrix}0&0&0\\1&0&0\\0&0&0\end{pmatrix}, \quad
  \tens{Y}_2=\begin{pmatrix}0&0&0\\0&0&0\\0&1&0\end{pmatrix}, \quad
  \tens{Y}_3=\begin{pmatrix}0&0&0\\0&0&0\\1&0&0\end{pmatrix}, \quad
  \tens{H}_2=\frac{1}{3}\diag(1,1,-2).
\end{gather*}
We denote the set $\{\tens{X}_1,\,\tens{Y}_1,\,\tens{H}_1,\,\tens{X}_3,\,\tens{Y}_3,\,\tens{X}_2,\,\tens{Y}_2,\,\tens{H}_2\}$  by $\{\tens{Z}_a\,{:}\,a\,{=}\,1,\,\dots,\, 8\}$.
With respect to the bilinear form $\langle \tens{A}, \tens{B} \rangle\,{=}\,\Tr \tens{A} \tens{B}$
we introduce the dual algebra
$\mathfrak{g}^{\ast}$ with the basis $\{\tens{Z}_a^{\ast}\}$:
$$\tens{X}_j^{\ast}=\tens{Y}_j,\ \ \tens{Y}_j^{\ast}=\tens{X}_j,\ j=1,\,2,\,3,\quad \tens{H}_1^{\ast}=\diag(1,-1,0),\ \
\tens{H}_2^{\ast}=\diag(0,1,-1).$$

Let $\mathcal{P}(\lambda,\lambda^{-1})$ be the algebra of
Laurent polynomials in $\lambda$, and
$\widetilde{\mathfrak{g}}$ be the loop algebra
$\mathfrak{sl}(3,\Complex)\otimes
\mathcal{P}(\lambda,\lambda^{-1})$. Then
\begin{equation*}
  \tens{Z}_a^{m} = \lambda^m \tens{Z}_a
\end{equation*}
is a \emph{basis} in $\widetilde{\mathfrak{g}}$. The loop algebra
is homogeneous or standard graded, this is easily checked by means of
the operator $\mathfrak{d} \,{=}\,d/d\lambda$ of homogeneous degree.
The superscript of $\tens{Z}_a^{m}$  indicates a homogeneous degree of
the basis element. By $\mathfrak{g}_m$, $m\in \Integer$ we denote the
eigenspace of homogeneous degree $m$, that is
\begin{equation*}
  \mathfrak{g}_{m}=\spanOp_{\Complex}
  \{\tens{X}_1^m,\,\tens{Y}_1^m,\,\tens{H}_1^m,\,\tens{X}_3^m,\,
  \tens{Y}_3^m,\,\tens{X}_2^m,\,\tens{Y}_2^m,\,\tens{H}_2^m\}.
\end{equation*}

According to the Kostant-Adler scheme \cite{AdlerMoerb}
$\widetilde{\mathfrak{g}}$ is decomposed into two subalgebras
\begin{equation*}
  \widetilde{\mathfrak{g}}_+ = \sum_{m\geqslant
  0}\mathfrak{g}_m,\qquad
  \widetilde{\mathfrak{g}}_- = \sum_{m < 0}\mathfrak{g}_m,\qquad
  \widetilde{\mathfrak{g}} = \widetilde{\mathfrak{g}}_+ +
  \widetilde{\mathfrak{g}}_-.
\end{equation*}
Further, we introduce the \emph{$\ad$-invariant bilinear forms}
\begin{equation*}
  \langle \tens{A}(\lambda), \tens{B}(\lambda) \rangle_{k} =
  \res_{\lambda=0} \lambda^{-k-1} \Tr \tens{A}(\lambda)
  \tens{B}(\lambda), \quad \tens{A}(\lambda),\ \tens{B}(\lambda)\in
  \widetilde{\mathfrak{g}}, \quad k\in \Integer
\end{equation*}
and use them to define the spaces dual to
$\widetilde{\mathfrak{g}}_+$ and $\widetilde{\mathfrak{g}}_-$.
\begin{example}\label{E:k-1}
Let $k=-1$. We have
\begin{equation*}
  (\widetilde{\mathfrak{g}}_-)^{\ast} = \widetilde{\mathfrak{g}}_+ ,\qquad
  (\widetilde{\mathfrak{g}}_+)^{\ast} = \widetilde{\mathfrak{g}}_-,
\end{equation*}
where $(\widetilde{\mathfrak{g}}_-)^{\ast}$ and
$(\widetilde{\mathfrak{g}}_+)^{\ast}$ contain only the nonzero
functionals on $\widetilde{\mathfrak{g}}_{\pm}$.
\end{example}
\begin{example}\label{E:kN}
Let $k=\mathcal{N}\,{-}\,1\geqslant 0$. Then
\begin{equation*}
  (\widetilde{\mathfrak{g}}_-)^{\ast} = \sum_{m\geqslant \mathcal{N}}
  \mathfrak{g}_{m},\qquad  (\widetilde{\mathfrak{g}}_+)^{\ast}=
  \sum_{m < \mathcal{N}} \mathfrak{g}_m.
\end{equation*}
\end{example}

\section{Orbits of $\mathfrak{sl}(3,\Complex)\otimes \mathcal{P}(\lambda,\lambda^{-1})$ as phase spaces}
Fixing $\mathcal{N}\,{\geqslant}\, 0$ we introduce the variables $\{\gamma_1^{(m)}$, $\beta_1^{(m)}$, $\alpha_1^{(m)}$, $\gamma_3^{(m)}$, $\beta_3^{(m)}$,
$\gamma_2^{(m)}$, $\beta_2^{(m)}$, $\alpha_2^{(m)}\,{:}$ $m\,{=}\,0$, $1$, $\dots$, $\mathcal{N}\}$
denoted all together by $\{L_a^{(m)}\,{:}$ $a\,{=}\,1$, \ldots, $8\}$. Consider the space
$\mathcal{M}\,{\in}\,\widetilde{\mathfrak{g}}^{\ast}$ of the elements
\begin{gather}\label{muExpr}
  \tens{L}(\lambda) =
  \sum_{m=0}^{\mathcal{N}} \sum_{a=1}^{\dim \mathfrak{g}} L_a^{(m)} \big(\tens{Z}_a^m\big)^\ast=  \begin{pmatrix}
  \alpha_1(\lambda) & \beta_1(\lambda) & \beta_3(\lambda) \\ \gamma_1(\lambda) &
  \alpha_2(\lambda)-\alpha_1(\lambda) & \beta_2(\lambda) \\ \gamma_3(\lambda) &
  \gamma_2(\lambda) & -\alpha_2(\lambda) \end{pmatrix},
\intertext{where} \nonumber
  \alpha_1(\lambda) = \sum_{m=0}^\mathcal{N} \lambda^m \alpha_{1}^{(m)},\quad
  \beta_1(\lambda) = \sum_{m=0}^{\mathcal{N}} \lambda^{m} \beta_{1}^{(m)},\quad
  \gamma_1(\lambda) = \sum_{m=0}^{\mathcal{N}} \lambda^{m} \gamma_{1}^{(m)},\\ \nonumber
  \alpha_2(\lambda) = \sum_{m=0}^\mathcal{N} \lambda^m \alpha_{2}^{(m)},\quad
  \beta_2(\lambda) = \sum_{m=0}^{\mathcal{N}} \lambda^{m} \beta_{2}^{(m)},\quad
  \gamma_2(\lambda) = \sum_{m=0}^{\mathcal{N}} \lambda^{m} \gamma_{2}^{(m)},\\ \nonumber
  \beta_3(\lambda) = \sum_{m=0}^{\mathcal{N}} \lambda^{m} \beta_{3}^{(m)},\quad
  \gamma_3(\lambda) = \sum_{m=0}^{\mathcal{N}} \lambda^{m} \gamma_{3}^{(m)}.
\end{gather}
Let $\mathcal{C}(\mathcal{M})$ be the space of smooth functions on $\mathcal{M}$. For
all $f_1$, $f_2\,{\in}\, \mathcal{C}(\mathcal{M})$ we define the \emph{first Lie-Poisson
bracket} by the formula
\begin{gather}\label{LiePoissonBraNLS}
  \{f_1,f_2\}_{\text{f}} = \sum_{m,n=0}^\mathcal{N} \sum_{a,b=1}^{\dim \mathfrak{g}} P_{ab}^{mn}(-1)
  \frac{\partial f_1}{\partial L_a^{(m)}}\frac{\partial f_2}{\partial L_b^{(n)}},\\
 P_{ab}^{mn}(-1) = \langle \tens{L}(\lambda),
  [\tens{Z}_a^{-m-1},\tens{Z}_b^{-n-1}] \rangle_{-1}. \nonumber
\end{gather}
The variables $\{L_a^{(\mathcal{N})}\}$ annihilate this Lie-Poisson bracket,
thus they are constant. To make the bracket nonsingular we restrict it to the subspace
$\mathcal{M}_0$ of $\mathcal{M}$ defined by the constraints
$$L_a^{(\mathcal{N})} = \const,\qquad a\,{=}\,1,\,\dots,\,8.$$
On $\mathcal{M}_0$ we define the \emph{second Lie-Poisson
bracket}  by the formula
\begin{gather}\label{LiePoissonBraHM}
  \{f_1,f_2\}_\text{s} = \sum_{m,n=0}^\mathcal{N} \sum_{a,b=1}^{\dim \mathfrak{g}} P_{ab}^{mn}(\mathcal{N}-1)
  \frac{\partial f_1}{\partial L_a^{(m)}}\frac{\partial
  f_2}{\partial L_b^{(n)}},  \\ P_{ab}^{mn}(\mathcal{N}-1) = \langle\tens{L}(\lambda),
   [\tens{Z}_a^{-m+\mathcal{N}-1},\tens{Z}_b^{-n+\mathcal{N}-1}] \rangle_{\mathcal{N}-1}.\notag
\end{gather}
\begin{remark}
In addition to the brackets (\ref{LiePoissonBraNLS}) and
(\ref{LiePoissonBraHM}), one can define intermediate brackets
with the Poisson tensors
\begin{equation}\label{LiePoissonBras}
 P_{ab}^{mn}(k)=\langle \tens{L}(\lambda),
   [\tens{Z}_a^{-m+k},\tens{Z}_b^{-n+k}] \rangle_{k}, \qquad k=0, \ldots, \mathcal{N}-2.
\end{equation}
\end{remark}

In what follows we consider the space $\mathcal{M}_0$, and use the set
$\{\gamma_{1}^{(m)}$, $\gamma_{2}^{(m)}$, $\gamma_{3}^{(m)}$, $\beta_{1}^{(m)}$,
$\beta_{2}^{(m)}$, $\beta_{3}^{(m)}$, $\alpha_{1}^{(m)}$, $\alpha_{2}^{(m)}\,{:}$ $m\,{=}\,0,\,
1,\, \dots,\, \mathcal{N}\,{-}\,1\}$ as \emph{dynamic variables} in it.
We call $\mathcal{M}_0$ a \emph{finite gap sector of~$\widetilde{\mathfrak{g}}$}, more precisely
the $\mathcal{N}$-gap sector. With respect to the bilinear form $\langle\cdot,\cdot \rangle_{-1}$ we introduce
the coadjoint action of $\widetilde{\mathfrak{g}}_-$ on $\mathcal{M}_0$. Indeed the
factor-algebra $\widetilde{\mathfrak{g}}_-/\sum_{l< {-}\mathcal{N}} \mathfrak{g}_l$
acts effectively on $\mathcal{M}_0$, that is $\mathcal{M}_0 \,{\subset}\,
(\widetilde{\mathfrak{g}}_-)^{\ast}$, see Example~\ref{E:k-1}.  The first Lie-Poisson bracket $\{\cdot,\cdot\}_{\text{f}}$
arises from the bilinear form $\langle\cdot,\cdot \rangle_{-1}$. On the other hand, we introduce
the coadjoint action of $\widetilde{\mathfrak{g}}_+$ on $\mathcal{M}_0$ with respect to the bilinear
form $\langle\cdot,\cdot \rangle_{\mathcal{N}-1}$: the factor-algebra
$\widetilde{\mathfrak{g}}_+/\sum_{l\geqslant \mathcal{N}} \mathfrak{g}_l$
acts effectively on $\mathcal{M}_0$, that is $\mathcal{M}_0\,{\subset}\, (\widetilde{\mathfrak{g}}_+)^{\ast}$,
see Example~\ref{E:kN}. The second Lie-Poisson bracket $\{\cdot,\cdot\}_{\text{s}}$ arises
from $\langle\cdot,\cdot\rangle_{\mathcal{N}-1}$.

Next, we introduce the $\ad^{\ast}$-invariant functions
\begin{align}
  &I_2(\lambda)\equiv \tfrac{1}{2} \Tr \tens{L}^2(\lambda) = \big[\alpha_1(\lambda)\big]^2 + \big[\alpha_2(\lambda)\big]^2 -
  \alpha_1(\lambda)\alpha_2(\lambda) + \beta_1(\lambda)\gamma_1(\lambda) + \notag
  \\ &\phantom{I_2(\lambda)\equiv \tfrac{1}{2} \Tr \widehat{L}^2(\lambda)} + \beta_2(\lambda)\gamma_2(\lambda) +
  \beta_3(\lambda)\gamma_3(\lambda) = \label{InvarF} \\ &\phantom{I_2(\lambda)} = \small
  -\begin{vmatrix} \alpha_1(\lambda) & \beta_1 (\lambda)  \\
  \gamma_1(\lambda) & \alpha_2(\lambda)-\alpha_1(\lambda) \end{vmatrix} -
  \begin{vmatrix} \alpha_2(\lambda)-\alpha_1(\lambda) & \beta_2 (\lambda) \\
  \gamma_2(\lambda) & -\alpha_2(\lambda) \end{vmatrix} -
  \begin{vmatrix} \alpha_1(\lambda) & \beta_3 (\lambda) \\
  \gamma_3(\lambda) & -\alpha_2(\lambda) \end{vmatrix}, \notag \\
&I_3(\lambda)\equiv\tfrac{1}{3} \Tr \tens{L}^3(\lambda) = \alpha_2(\lambda)\big[\alpha_1(\lambda)\big]^2 -
  \alpha_1(\lambda) \big[\alpha_2(\lambda)\big]^2 +
  \beta_1(\lambda)\gamma_1(\lambda)\alpha_2(\lambda) + \notag
  \\ &\phantom{I_2(\lambda)\equiv\tfrac{1}{3} \Tr \widehat{L}^3(\lambda)}
  + \beta_2(\lambda)\big[\beta_1(\lambda)\gamma_3(\lambda)-\gamma_2(\lambda)\alpha_1(\lambda)\big] + \notag
  \\ &\phantom{I_2(\lambda)\equiv\tfrac{1}{3} \Tr \widehat{L}^3(\lambda)} + \beta_3(\lambda)\big[
  \gamma_1(\lambda)\gamma_2(\lambda)-[\alpha_2(\lambda)-\alpha_1(\lambda)]\gamma_3(\lambda)\big]
  = \det \tens{L}(\lambda). \notag
\end{align}
Every function $I_k$ is a sum of the diagonal $k^{\text{th}}$ minors with an accuracy of the sign. The functions $I_2$, $I_3$ are polynomials in the spectral parameter $\lambda$,
and their coefficients serve as invariant functions in dynamic variables, namely:
\begin{align}
  &\quad I_2(\lambda)= h_0 + h_1 \lambda + \cdots + h_{2\mathcal{N}}\lambda^{2\mathcal{N}},\quad
  I_3(\lambda)= f_0 + f_1 \lambda + \cdots + f_{3\mathcal{N}}\lambda^{3\mathcal{N}}, \notag \\
  &h_{\nu} = \small -\sum_{m+n=\nu} \left( \begin{vmatrix} \alpha_1^{(m)} & \beta_1^{(n)}  \\
  \gamma_1^{(m)} & \alpha_2^{(n)}-\alpha_1^{(n)} \end{vmatrix} +
  \begin{vmatrix} \alpha_2^{(m)}-\alpha_1^{(m)} & \beta_2^{(n)} \\
  \gamma_2^{(m)} & -\alpha_2^{(n)} \end{vmatrix} +
  \begin{vmatrix} \alpha_1^{(m)} & \beta_3^{(n)} \\
  \gamma_3^{(m)} & -\alpha_2^{(n)} \end{vmatrix} \right), \notag \\
  &\phantom{I_2(\lambda)} \nu=0,\,1,\,\dots,\,2\mathcal{N}; \label{InvFunc}  \\
  &f_{\nu} = \small \sum_{m+n+k=\nu} \begin{vmatrix}
  \alpha_1^{(m)} & \beta_1^{(n)} & \beta_3^{(k)} \\ \gamma_1^{(m)} &
  \alpha_2^{(n)}-\alpha_1^{(n)} & \beta_2^{(k)} \\ \gamma_3^{(m)} &
  \gamma_2^{(n)} & -\alpha_2^{(k)} \end{vmatrix},
  \quad \nu=0,\,1,\,\dots,\,3\mathcal{N}. \notag
\end{align}
Evidently, $h_{2\mathcal{N}}$, $f_{3\mathcal{N}}$ are constant, for they do not contain dynamic variables.

The following assertions are immediately derived from the
Kostant-Adler scheme \cite{AdlerMoerb}.
\begin{proposition}
All functions $h_{\nu}$, $\nu\,{=}\,0,\,1,\,\dots,\,2\mathcal{N}\,{-}\,1$ and
$f_{\nu}$, $\nu\,{=}\,0$, $1$, $\dots$, $3\mathcal{N}\,{-}\,1$ defined by
\eqref{InvFunc} are in involution with respect to the brackets
\eqref{LiePoissonBraNLS} and \eqref{LiePoissonBraHM}.
That is, these functions serve as integrals of motion.
\end{proposition}
\begin{proposition}\label{P:BraNLSannihil}
The functions $h_{\nu}$, $f_{\nu+\mathcal{N}}$, $\nu\,{=}\,\mathcal{N},\,\dots,\, 2\mathcal{N}\,{-}\,1$
are functionally independent on
$\mathcal{M}_0$ and annihilate the first Lie-Poisson bracket \eqref{LiePoissonBraNLS},
namely they are Casimir functions with respect to the first Lie-Poisson bracket.
The rest of integrals of motion: $h_{\nu}$, $\nu\,{=}\,0,\,\dots,\, \mathcal{N}\,{-}\,1$,
and $f_{\nu}$, $\nu\,{=}\,0,\,\dots,\, 2\mathcal{N}\,{-}\,1$
serve as Hamiltonians with respect to the first Lie-Poisson bracket.
\end{proposition}
Let $\mathcal{O}_{\text{f}}\,{\subset}\,
\mathcal{M}_0$ be the algebraic manifold defined by
\begin{equation}\label{OrbEqNLS}
h_{\nu}\,{=}\,c_{\nu},\quad
f_{\nu+\mathcal{N}}\,{=}\,d_{\nu+\mathcal{N}},\qquad
\nu\,{=}\,\mathcal{N},\, \ldots,\, 2\mathcal{N}\,{-}\,1,
\end{equation}
where all $c_{\nu}$, $d_{\nu+\mathcal{N}}$
are fixed complex numbers. The manifold $\mathcal{O}_{\text{f}}$
is a generic orbit of coadjoint action of the
subalgebra $\widetilde{\mathfrak{g}}_{-}$ on $\mathcal{M}_0$,
$\dim \mathcal{O}_{\text{f}}\,{=}\,6\mathcal{N}$.
Variation of the constants $c_{\nu}$, $d_{\nu+\mathcal{N}}$
gives a foliation of $\mathcal{M}_0$ into orbits of the first type.
Every orbit serves as a symplectic leaf in the symplectic manifold $\mathcal{M}_0$.

\begin{proposition}\label{P:BraHMannihil}
The functions $h_{\nu}$, $f_{\nu}$, $\nu\,{=}\,0,\,\dots,\, \mathcal{N}\,{-}\,1$ are functionally
independent on $\mathcal{M}_0$ and annihilate the second Lie-Poisson bracket~\eqref{LiePoissonBraHM},
namely they are Casimir functions with respect to the second Lie-Poisson bracket.
The rest of integrals of motion: $h_{\nu}$, $\nu\,{=}\,\mathcal{N},\,\dots,\, 2\mathcal{N}\,{-}\,1$,
and $f_{\nu}$, $\nu\,{=}\,\mathcal{N},\,\dots,\, 3\mathcal{N}\,{-}\,1$
serve as Hamiltonians with respect to the second Lie-Poisson bracket.
\end{proposition}
The algebraic manifold $\mathcal{O}_{\text{s}}\,{\subset}\,\mathcal{M}_0$ defined by
\begin{equation}\label{OrbEqHM}
h_{\nu}\,{=}\,c_{\nu},\quad f_{\nu}\,{=}\,d_{\nu},\qquad
\nu=0,\, \dots,\, \mathcal{N}\,{-}\,1,
\end{equation}
where all $c_{\nu}$, $d_{\nu}$ are fixed complex numbers,
is a generic orbit of coadjoint action of the
subalgebra~$\widetilde{\mathfrak{g}}_{+}$ on $\mathcal{M}_0$,
$\dim \mathcal{O}_{\text{s}}\,{=}\,6\mathcal{N}$.
Variation of the constants $c_{\nu}$, $d_{\nu}$
gives another foliation of $\mathcal{M}_0$ into orbits of the second type.
In what follows we call $\mathcal{O}_{\text{f}}$ and $\mathcal{O}_{\text{s}}$ simply \emph{orbits},
and call \eqref{OrbEqNLS}, \eqref{OrbEqHM} orbit equations.

\section{Separation of variables on $\mathcal{O}_\text{f}$}
The orbit $\mathcal{O}_\text{f}$ with the first Lie-Poisson bracket \eqref{LiePoissonBraNLS}
has the following Poisson structure:
\begin{gather}\label{PoissonBraNLS}
\{L_{ij}^{(m)}, L_{kl}^{(n)}\}_\text{f} = L_{kj}^{(m+n+1)}\delta_{il} - L_{il}^{(m+n+1)} \delta_{kj},
\end{gather}
also written in terms of the $\tens{r}$-matrix
\begin{gather}
\tens{r}_{12}(u-v) = \frac{1}{u-v}\, \sum_{a,b} \langle \tens{Z}_a,
\tens{Z}_b \rangle \tens{Z}_a^{\ast} \otimes \tens{Z}_b^{\ast}
\label{Rmatrix}\\
\{\tens{L}_1(u)\overset{\otimes}{,}\, \tens{L}_2(v)\}_\text{f} =
  [\tens{r}_{12}(u-v),\tens{L}_1(u)+\tens{L}_2(v)] \nonumber
\end{gather}
with $\tens{L}_1(u)\,{=}\,\tens{L}(u)\,{\otimes}\, \Ibb$,
$\tens{L}_2(v)\,{=}\,\Ibb \,{\otimes}\, \tens{L}(v)$, where $\Ibb$ is the identity matrix.

We parameterize the orbit $\mathcal{O}_\text{f}$ by the dynamic
variables $\{\gamma_1^{(m)}$, $\gamma_2^{(m)}$, $\gamma_3^{(m)}$,
$\beta_1^{(m)}$, $\alpha_{1}^{(m)}$,  $\alpha_{2}^{(m)}{:}$ $m\,{=}\,0,\,1$, \ldots, $\mathcal{N}\,{-}\,1\}$, that
is we eliminate the set $\{\beta_{2}^{(m)}, \beta_{3}^{(m)}\}$. One can choose another set to eliminate,
requiring all the invariant functions are linear in this set of variables. Thus, the other possible sets of eliminated
variables are the following: $\{\beta_{1}^{(m)}, \beta_{3}^{(m)}\}$, $\{\gamma_{2}^{(m)}, \gamma_{3}^{(m)}\}$, or
$\{\gamma_{1}^{(m)}, \gamma_{3}^{(m)}\}$.
Using linearity of the orbit equations \eqref{OrbEqNLS} in the variables
$\{\beta_{2}^{(m)}, \beta_{3}^{(m)}{:}\,m\,{=}\,0,\,1$, \ldots, $\mathcal{N}\}$
one can write them in the matrix form
\begin{subequations}
\begin{gather} \label{ConstrNLS}
\vec{c}_\text{f} = \tens{F}^+ \vec{\beta} + \vec{\eta}^{+}_\text{f}, \\
\nonumber
\tens{F}^+ =\small \begin{bmatrix}
    \tens{F}_{\mathcal{N}} & \tens{F}_{\mathcal{N}-1} & \dots & \tens{F}_{1} & \tens{F}_{0} \\
   0 & \tens{F}_{\mathcal{N}} & \dots & \tens{F}_{2} & \tens{F}_{1} \\
   \vdots & \vdots& \ddots& \vdots& \vdots \\
   0 & 0 & \dots & \tens{F}_{\mathcal{N}} & \tens{F}_{\mathcal{N}-1} \\
   0 & 0 & \dots & 0 &  \tens{F}_{\mathcal{N}}
   \end{bmatrix},\quad
\vec{\beta} = \begin{bmatrix} \vec{\beta}^{(0)} \\ \vec{\beta}^{(1)} \\ \vdots \\
   \vec{\beta}^{(\mathcal{N}-1)} \\ \vec{\beta}^{(\mathcal{N})}  \end{bmatrix}, \quad
\vec{c}_\text{f} = \begin{bmatrix} \vec{c}_{\mathcal{N}}^{\text{f}} \\ \vec{c}_{\mathcal{N}+1}^{\text{f}} \\ \vdots \\
   \vec{c}_{2\mathcal{N}-1}^{\text{f}} \\ \vec{c}_{2\mathcal{N}}^{\text{f}} \end{bmatrix}, \quad
\vec{\eta}^{+}_\text{f} = \begin{bmatrix} \vec{\eta}_{\mathcal{N}}^{\text{f}} \\
\vec{\eta}_{\mathcal{N}+1}^{\text{f}} \\ \vdots \\
\vec{\eta}_{2\mathcal{N}-1}^{\text{f}} \\ \vec{\eta}_{2\mathcal{N}}^{\text{f}} \end{bmatrix},\\
\tens{F}_{j} = \small \begin{bmatrix} \gamma_2^{(j)} &  \gamma_3^{(j)} \\ \Gamma_2^{(j+\mathcal{N})}  & \Gamma_3^{(j+\mathcal{N})} \end{bmatrix},\quad
\vec{\beta}^{(j)} = \begin{bmatrix} \beta_2^{(j)} \\ \beta_3^{(j)} \end{bmatrix}, \quad
\vec{c}_{j}^{\text{f}} = \begin{bmatrix} c_{j} \\ d_{j+\mathcal{N}} \end{bmatrix}, \quad
\vec{\eta}_{j}^{\text{f}} = \begin{bmatrix} \eta_{j} \\ H_{j+\mathcal{N}}  \end{bmatrix},\\
\eta_{j} = -A^{(j)} + \sum\limits_{m+n=j} \alpha_2^{(m)}\alpha_2^{(n)},\quad
H_{j} = -\sum\limits_{m+n=j} \alpha_2^{(m)} A^{(n)}, \label{HetaNLS} \\
\label{GammaDef}
\Gamma_2^{(j)} = \small -\sum\limits_{m+n=j}
   \begin{vmatrix} \alpha_1^{(m)} & \beta_1^{(n)}  \\
  \gamma_3^{(m)} & \gamma_2^{(n)} \end{vmatrix},\quad
\Gamma_3^{(j)} =  \sum\limits_{m+n=j}
    \begin{vmatrix} \gamma_1^{(m)} & \alpha_2^{(n)} - \alpha_1^{(n)}   \\
  \gamma_3^{(m)} & \gamma_2^{(n)} \end{vmatrix},  \\ \label{ADef}
A^{(j)} =  \small \sum\limits_{m+n=j}
    \begin{vmatrix} \alpha_1^{(m)} & \beta_1^{(n)}  \\
  \gamma_1^{(m)} & \alpha_2^{(n)}-\alpha_1^{(n)} \end{vmatrix}.
\end{gather}
\end{subequations}
Supposing $\tens{F}_{\mathcal{N}}$ is nonsingular, one easily eliminates the variables $\vec{\beta}$
\begin{gather*}
\vec{\beta} = (\tens{F}^+)^{-1} (\vec{c}_\text{f} - \vec{\eta}^{+}_\text{f}), \qquad
\text{or} \\ \small
\begin{bmatrix} \vec{\beta}_{0} \\ \vec{\beta}_{1} \\ \vdots \\
   \vec{\beta}_{\mathcal{N}-1} \\ \vec{\beta}_{\mathcal{N}}  \end{bmatrix} =
   \begin{bmatrix}
   \tens{F}_{\mathcal{N}}^{-1} & \widetilde{\tens{F}}_{\mathcal{N}-1} & \dots & \widetilde{\tens{F}}_{1} & \widetilde{\tens{F}}_{0}\\
   0 & \tens{F}_{\mathcal{N}}^{-1} & \dots & \widetilde{\tens{F}}_{2} & \widetilde{\tens{F}}_{1} \\
   \vdots & \vdots& \ddots& \vdots& \vdots \\
   0 & 0 & \dots & \tens{F}_{\mathcal{N}}^{-1} & \widetilde{\tens{F}}_{\mathcal{N}-1} \\
   0 & 0 & \dots & 0 & \tens{F}_{\mathcal{N}}^{-1}
   \end{bmatrix} \begin{bmatrix}
   \vec{c}_{\mathcal{N}}^{\text{f}} - \vec{\eta}_{\mathcal{N}}^{\text{f}} \\
   \vec{c}_{\mathcal{N}+1}^{\text{f}} - \vec{\eta}_{\mathcal{N}+1}^{\text{f}} \\ \vdots \\
   \vec{c}_{2\mathcal{N}-1}^{\text{f}} - \vec{\eta}_{2\mathcal{N}-1}^{\text{f}} \\
   \vec{c}_{2\mathcal{N}}^{\text{f}} - \vec{\eta}_{2\mathcal{N}}^{\text{f}} \end{bmatrix},\\
\widetilde{\tens{F}}_{\mathcal{N}-n} = \tens{F}_{\mathcal{N}}^{-1}
\sum_{k=1}^n \big(-\tens{F}_{\mathcal{N}-n-1+k} \tens{F}_{\mathcal{N}}^{-1}\big)^k, \qquad n=1,\,\dots,\,\mathcal{N}.
\end{gather*}
Next, substitute $\vec{\beta}$ into  the Hamiltonians $h_{0}, h_{1}$, $\ldots$,
$h_{\mathcal{N}-1}$, $f_0$, $f_1$, $\ldots$, $f_{2\mathcal{N}-1}$
\begin{equation}\label{HamiltNLS}
   \vec{h}_\text{f} = \tens{F}^{-} \vec{\beta} +  \vec{\eta}^{-}_\text{f} =
   \tens{F}^{-} (\tens{F}^{+})^{-1} \vec{c}_\text{f} + \vec{\eta}^{-}_\text{f} -
   \tens{F}^{-} (\tens{F}^{+})^{-1} \vec{\eta}^{+}_\text{f},
\end{equation}
where
\begin{gather*}
\small
   \tens{F}^{-} = \begin{bmatrix}
   \tens{g}_0 & 0 & \dots & 0 & 0 \\
   \tens{g}_1 & \tens{g}_0 & \dots & 0 & 0 \\
   \vdots & \vdots& \ddots& \vdots& \vdots \\
   \tens{g}_{\mathcal{N}-1} & \tens{g}_{\mathcal{N}-2} & \dots & \tens{g}_0 & 0 \\
   \tens{F}_{0} & \tens{g}_{\mathcal{N}-1}^0 & \dots & \tens{g}_1^0 & \tens{g}_0^0 \\
   \tens{F}_{1} & \tens{F}_{0} & \dots & \tens{g}_2^0 & \tens{g}_1^0 \\
   \vdots & \vdots& \ddots& \vdots& \vdots \\
   \tens{F}_{\mathcal{N}-1} & \tens{F}_{\mathcal{N}-2} & \dots &  \tens{F}_{0} & \tens{g}_{\mathcal{N}-1}^0
   \end{bmatrix},\quad
\vec{h}_\text{f} = \begin{bmatrix} f_0 \\ f_1 \\ \vdots \\  f_{\mathcal{N}-1} \\
   \vec{h}_0^{\text{f}} \\ \vec{h}_1^{\text{f}} \\  \vdots \\ \vec{h}_{\mathcal{N}-1}^{\text{f}}
   \end{bmatrix},   \quad
\vec{\eta}^{-}_\text{f} = \begin{bmatrix} H_0 \\ H_1 \\ \vdots \\ H_{\mathcal{N}-1}
    \\ \vec{\eta}_0^{\text{f}} \\ \vec{\eta}_1^{\text{f}} \\  \vdots \\ \vec{\eta}_{\mathcal{N}-1}^{\text{f}}  \end{bmatrix},\\
\small
\tens{g}_j = \begin{bmatrix} \Gamma_2^{(j)}  & \Gamma_3^{(j)}  \end{bmatrix}, \quad
\tens{g}_j^0 = \begin{bmatrix} 0 & 0 \\ \Gamma_2^{(j)}  & \Gamma_3^{(j)}  \end{bmatrix}, \quad
\vec{h}_j^{\text{f}} = \begin{bmatrix} h_{j} \\ f_{j+\mathcal{N}} \end{bmatrix}.
\end{gather*}
Note that the expressions (\ref{HamiltNLS}) are linear in
$\{c_{\nu},\,d_{\nu+\mathcal{N}}\,{:}\, \nu=\mathcal{N}, \ldots, 2\mathcal{N}\}$.

To proceed we need to define the \emph{characteristic polynomial}
\begin{gather}\label{CharPoly}
  P(w,\lambda) = \det\bigl(\tens{L}(\lambda)- w \Ibb \bigr).
\end{gather}
It defines the spectral curve~$\mathcal{R}$:
\begin{equation}\label{SpectrCurve}
  w^3 - I_2(\lambda) w - I_3(\lambda) = 0,
\end{equation}
which is a curve of genus $3\mathcal{N}\,{-}2$ in general.
The spectral curve is common for
integrable systems on orbits of both the types: $\mathcal{O}_\text{f}$ and $\mathcal{O}_\text{s}$.
Restriction to an orbit is realized by implementation of the orbit equations \eqref{OrbEqNLS}
or \eqref{OrbEqHM}, which fix some coefficients in \eqref{SpectrCurve}. The rest of coefficients
serve as Hamiltonians on the orbit and also remain constant during the evolution of a system.

Consider the spectral curve restricted to the orbit $\mathcal{O}_\text{f}$.
Denoting its points by $\{(\lambda_k,\,
w_k)\}$ we write the following set of equations
\begin{multline}\label{HyperCurveNLS}
  w_k^3 = w_k\Big(h_{0}+h_1 \lambda_k + \cdots h_{\mathcal{N}-1}\lambda_k^{\mathcal{N}-1} +
  c_{\mathcal{N}} \lambda_k^{\mathcal{N}} + c_{\mathcal{N}+1} \lambda_k^{\mathcal{N}+1} + \cdots + c_{2\mathcal{N}} \lambda_k^{2\mathcal{N}}\Big)
  + \\ +\Big(f_{0} + f_1 \lambda_k + \cdots f_{2\mathcal{N}-1}\lambda_k^{2\mathcal{N}-1} +
  d_{2\mathcal{N}} \lambda_k^{2\mathcal{N}} + d_{2\mathcal{N}+1} \lambda_k^{2\mathcal{N}+1} + \cdots + d_{3\mathcal{N}} \lambda_k^{3\mathcal{N}}\Big),
\end{multline}
$k\,{=}\,1$, $\dots$, $3\mathcal{N}$,
or in the matrix form
\begin{gather*}
\tens{W}^{-}_\text{f} \vec{h}_\text{f} +  \tens{W}^{+}_\text{f} \vec{c}_\text{f} = \vec{w}^{\text{cubed}}, \\
\small
\tens{W}^{-}_\text{f} =  \begin{bmatrix}
1 & \lambda_1 & \dots & \lambda_1^{\mathcal{N}-1} & \tens{W}_1 & \lambda_1 \tens{W}_1 & \dots & \lambda_1^{\mathcal{N}-1} \tens{W}_1 \\
1 & \lambda_2 & \dots & \lambda_2^{\mathcal{N}-1} & \tens{W}_2 & \lambda_2 \tens{W}_2 & \dots & \lambda_2^{\mathcal{N}-1} \tens{W}_2 \\
\vdots & \vdots & \dots & \vdots & \vdots & \vdots & \dots & \vdots \\
1 & \lambda_{3\mathcal{N}} & \dots & \lambda_{3\mathcal{N}}^{\mathcal{N}-1} &
\tens{W}_{3\mathcal{N}} & \lambda_{3\mathcal{N}} \tens{W}_{3\mathcal{N}} & \dots &
\lambda_{3\mathcal{N}}^{\mathcal{N}-1} \tens{W}_{3\mathcal{N}} \end{bmatrix},\qquad
\tens{W}_k \,{=}\, \begin{bmatrix} w_k & \lambda_k^{\mathcal{N}} \end{bmatrix},\\
\tens{W}^{+}_\text{f} = \small \begin{bmatrix}
\lambda_1^{\mathcal{N}} \tens{W}_1 & \lambda_1^{\mathcal{N}+1} \tens{W}_1 & \dots & \lambda_1^{2\mathcal{N}} \tens{W}_1 \\
\lambda_2^{\mathcal{N}} \tens{W}_2 & \lambda_2^{\mathcal{N}+1} \tens{W}_2 & \dots & \lambda_2^{2\mathcal{N}} \tens{W}_2 \\
\vdots & \vdots & \dots & \vdots \\
\lambda_{3\mathcal{N}}^{\mathcal{N}} \tens{W}_{3\mathcal{N}} & \lambda_{3\mathcal{N}}^{\mathcal{N}+1} \tens{W}_{3\mathcal{N}}
& \dots & \lambda_{3\mathcal{N}}^{2\mathcal{N}} \tens{W}_{3\mathcal{N}}  \end{bmatrix},\qquad
\vec{w}^{\text{cubed}} = \begin{bmatrix} w_1^3 \\ w_2^3 \\ \vdots \\ w_{3\mathcal{N}}^3  \end{bmatrix}.
\end{gather*}
and solve them for $3\mathcal{N}$ Hamiltonians.
Suppose that all pairs $\{(\lambda_k,\, w_k)\,{:}$ $k\,{=}\,1$,
\ldots, $3\mathcal{N}\}$ are distinct points and $\tens{W}^-_{\text{f}}$ is nonsingular,
then the Hamiltonians can be expressed by the formula
\begin{equation}\label{hSystNLS}
 \vec{h}_\text{f}  =  - (\tens{W}^{-}_\text{f})^{-1} \tens{W}^{+}_\text{f} \vec{c}_\text{f} + (\tens{W}^{-}_\text{f})^{-1}\vec{w}^{\text{cubed}}.
\end{equation}
On the orbit $\mathcal{O}_\text{f}$ the formulas (\ref{HamiltNLS}) and (\ref{hSystNLS})
define the same set of functions, moreover, both
of them are linear in $\{c_{\nu},\,d_{\nu+\mathcal{N}}\,{:}$
$\nu\,{=}\,\mathcal{N}$, \ldots, $2\mathcal{N}\}$. As $\{c_{\nu},\,d_{\nu+\mathcal{N}}\}$ are
independent parameters one can equate the corresponding
terms, that is
\begin{gather}
 \tens{F}^{-} (\tens{F}^{+})^{-1} = - (\tens{W}^{-}_\text{f})^{-1} \tens{W}^{+}_\text{f},\qquad
 \vec{\eta}^{-}_\text{f} - \tens{F}^{-} (\tens{F}^{+})^{-1} \vec{\eta}^{+}_\text{f} =
 (\tens{W}^{-}_\text{f})^{-1}\vec{w}^{\text{cubed}}\quad \Rightarrow \nonumber\\
 \tens{W}^{-}_\text{f} \tens{F}^{-}  + \tens{W}^{+}_\text{f} \tens{F}^{+} = 0,\qquad
 \tens{W}^{-}_\text{f} \vec{\eta}^{-}_\text{f} +
 \tens{W}^{+}_\text{f} \vec{\eta}^{+}_\text{f} = \vec{w}^{\text{cubed}}. \label{RootEqNLS}
\end{gather}
The first matrix equation \eqref{RootEqNLS} gives the following
\begin{subequations}\label{GammaEqNLS}
\begin{gather}\label{Gamma2EqNLS}
\Gamma_2^{(0)} + \Gamma_2^{(1)} \lambda + \cdots + \Gamma_2^{(2\mathcal{N})} \lambda_k^{2\mathcal{N}}
+ w_k \big(\gamma_2^{(0)} + \cdots + \gamma_2^{(\mathcal{N})} \lambda_k^{\mathcal{N}}\big) = 0,\\
\label{Gamma3EqNLS}
\Gamma_3^{(0)} + \Gamma_3^{(1)}\lambda + \cdots + \Gamma_3^{(2\mathcal{N})} \lambda_k^{2\mathcal{N}}
+ w_k \big(\gamma_3^{(0)} + \cdots + \gamma_3^{(\mathcal{N})} \lambda_k^{\mathcal{N}}\big) = 0
\end{gather}
or more concisely
\begin{equation}\label{Minors23}
\Gamma_2(\lambda_k) + w_k \gamma_2(\lambda_k) = 0,\qquad
\Gamma_3(\lambda_k) + w_k \gamma_3(\lambda_k) = 0,
\end{equation}
\end{subequations}
where $\Gamma_2$ and $\Gamma_3$ are polynomials of degree $2\mathcal{N}$ in general,
and at least $\Gamma_2^{(2\mathcal{N})}$, $\gamma_3^{(\mathcal{N})}$ or
$\Gamma_3^{(2\mathcal{N})}$, $\gamma_2^{(\mathcal{N})}$ are nonzero.
The $6\mathcal{N}$ equations \eqref{Minors23} are consistent if
\begin{equation}\label{ConsistentEqNLS}
\begin{vmatrix} \gamma_2 (\lambda_k) & \gamma_3 (\lambda_k)\\
\Gamma_2(\lambda_k) & \Gamma_3(\lambda_k) \end{vmatrix} =0.
\end{equation}
The second matrix equation \eqref{RootEqNLS} gives
\begin{gather}\label{SpectCurveNLS}
w_k^3 = -\alpha_2(\lambda_k) A(\lambda_k) + w_k \big(\alpha_2^2(\lambda_k)-A(\lambda_k)\big),\quad \text{or}\\
\label{SpectralCurveRed}
\big[w_k + \alpha_2(\lambda_k)\big]  \big[w_k^2 - \alpha_2(\lambda_k) w_k + A(\lambda_k) \big] =0, \nonumber
\end{gather}
which is a simplification of the spectral curve equation \eqref{SpectrCurve}
realized at every point $(\lambda_k,w_k)$. Here
\begin{gather*}
A(\lambda)\,{=}\,A^{(0)} + A^{(1)}\lambda + \cdots + A^{(2\mathcal{N})}\lambda^{2\mathcal{N}}.
\end{gather*}

The set $\{\vec{\lambda},\,\vec{w}\}\,{\equiv}\,\{(\lambda_k,\, w_k)\,{:}$ $k\,{=}\,1,\,
\dots,\, 3\mathcal{N}\}$ of variables defined by \eqref{ConsistentEqNLS}, \eqref{Minors23}
are points of the spectral curve $\mathcal{R}$ obtained from its restriction \eqref{HyperCurveNLS}
to the orbit $\mathcal{O}_{\text{f}}$.
The variables $\{\vec{\lambda},\,\vec{w}\}$ give another parametrization of $\mathcal{O}_\text{f}$,
we call them \emph{spectral variables}, they serve as \emph{variables of separation} as shown below.

Now we trace the sequence of changes of variables on the orbit. First of all
it is suitable to change the dynamic variables $\{\gamma_1^{(m)}$, $\beta_1^{(m)}$,
$\alpha_1^{(m)}$, $\gamma_3^{(m)}$, $\gamma_2^{(m)}$,
 $\alpha_2^{(m)}\,{:}$ $m\,{=}\,0,\,1,\,\dots,\,\mathcal{N}\,{-}\,1\}$ into the following:
$\{\vec{\gamma}_2$, $\vec{\Gamma}_2$, $\vec{\alpha}_2$, $\vec{A}\}$ that is
\begin{gather*}
\Big\{ \gamma_2^{(0)},\, \gamma_2^{(1)},\,\dots,\, \gamma_2^{(\mathcal{N}-1)},\,
\Gamma_2^{(0)},\, \Gamma_2^{(1)},\, \dots,\, \Gamma_2^{(2\mathcal{N}-1)},  \\
\alpha_2^{(0)},\, \alpha_2^{(1)},\,
\dots, \, \alpha_2^{(\mathcal{N}-1)},\,A^{(0)},\, A^{(1)},\, \dots, \, A^{(2\mathcal{N}-1)} \Big\},
\end{gather*}
according to \eqref{GammaDef}, \eqref{ADef}. Equally one can use the set
\begin{gather*}
\Big\{ \gamma_3^{(0)},\, \gamma_3^{(1)},\,\dots,\, \gamma_3^{(\mathcal{N}-1)},\,
\Gamma_3^{(0)},\, \Gamma_3^{(1)},\, \dots,\, \Gamma_3^{(2\mathcal{N}-1)}, \\
\alpha_2^{(0)},\, \alpha_2^{(1)},\,
\dots, \, \alpha_2^{(\mathcal{N}-1)},\,A^{(0)},\, A^{(1)},\, \dots, \, A^{(2\mathcal{N}-1)} \Big\}.
\end{gather*}
Then from \eqref{HamiltNLS} the variables $\{\vec{\alpha}_2,\,\vec{A}\}$ are replaced by
the Hamiltonians $\vec{h}_\text{f}\,{=}\,\{h_0$, $h_1$, \ldots, $h_{\mathcal{N}-1}$, $f_0$, $f_1$, \ldots, $f_{2\mathcal{N}-1}\}$.
In this way $\{\vec{\gamma}_2$, $\vec{\Gamma}_2$, $\vec{\alpha}_2,\,\vec{A}\}$ is changed into
$\{\vec{\gamma}_2$, $\vec{\Gamma}_2$, $\vec{h}_\text{f}\}$. At last \eqref{Gamma2EqNLS}, \eqref{HyperCurveNLS}
connect the latter to the spectral variables $\{\vec{\lambda},\,\vec{w}\}$.
The equations \eqref{Minors23} give direct relations between the dynamic and the spectral variables;
they are used for computation.

Using the conventional notations we formulate Conjecture\;1 from \cite{Sklyanin92} for our purpose
\begin{SoV}\label{T:SoVNLS}
Suppose the orbit $\mathcal{O}_{\text{f}}$ is parameterized by the variables
 $\{\gamma_1^{(m)}$, $\beta_1^{(m)}$, $\alpha_1^{(m)}$, $\gamma_3^{(m)}$, $\gamma_2^{(m)}$,
 $\alpha_2^{(m)}\,{:}$ $m\,{=}\,0,\,\dots,\,\mathcal{N}\,{-}\,1\}$ as above.
Then the new variables $\{(\lambda_k, w_k)\,{:}$ $k\,{=}\,1,\,\dots,\,3\mathcal{N}\}$
defined by the formulas
\begin{equation}\label{newvarNLS}
  \mathcal{B}(\lambda_k)=0,\qquad
  w_k = \mathcal{A}(\lambda_k),
\end{equation}
where $\mathcal{B}$ is the polynomial of degree $3\mathcal{N}$ and $\mathcal{A}$
is the algebraic function such that
\begin{subequations}\label{ABNLS}
\begin{align}
\mathcal{B}(\lambda) &= \big[2\alpha_1(\lambda)-\alpha_2(\lambda)\big] \gamma_2(\lambda)
  \gamma_3(\lambda) - \beta_1(\lambda) \gamma_3^2(\lambda) + \gamma_1(\lambda)\gamma_2^2(\lambda)
  = \begin{vmatrix} \gamma_2(\lambda) & \gamma_3 (\lambda) \\ \Gamma_2(\lambda) & \Gamma_3 (\lambda) \end{vmatrix},
  \label{NilpotentPolyNLS}\\   \label{w1PolyNLS}
  \mathcal{A}(\lambda) & = \alpha_1(\lambda) -\frac{\beta_1(\lambda) \gamma_3(\lambda)}{\gamma_2(\lambda)}
  \equiv - \frac{\Gamma_2(\lambda)}{\gamma_2(\lambda)} \quad \text{or} \\
\label{w2PolyNLS}  & = \alpha_2(\lambda) - \alpha_1(\lambda) -
  \frac{\gamma_1(\lambda) \gamma_2(\lambda)}{\gamma_3(\lambda)} \equiv
  -\frac{\Gamma_3(\lambda)}{\gamma_3(\lambda)}
\end{align}
\end{subequations}
have the following properties:
\begin{enumerate}
\item[\textup{(i)}]  a pair $(\lambda_k,w_k)$ is a root of the characteristic
polynomial \eqref{CharPoly};
\item[\textup{(ii)}] a pair
$(\lambda_k, w_k)$ is canonically conjugate with respect to
the first Lie-Poisson bracket \eqref{LiePoissonBraNLS}:
\begin{equation}\label{PoissonBraNLSConj}
  \{\lambda_k,\lambda_l\}_{\textup{f}} =0, \qquad
  \{\lambda_k, w_l\}_{\textup{f}} = \delta_{kl}, \qquad
  \{w_k,w_l\}_{\textup{f}}=0;
\end{equation}
\item[\textup{(iii)}] the corresponding  Liouville 1-form is
\begin{gather*}
\Omega_{\textup{f}}=\sum\limits_{k} w_{k}\,d\lambda_{k}.
\end{gather*}
\end{enumerate}
\end{SoV}

\begin{proof}
(i) The equation \eqref{NilpotentPolyNLS} is equivalent to the consistent condition \eqref{ConsistentEqNLS}. This implies
that two expressions \eqref{w1PolyNLS}, \eqref{w2PolyNLS} for $\mathcal{A}$ coincide at all zeros
$\{\lambda_k\}$ of $\mathcal{B}$. The characteristic polynomial $P$ defined by \eqref{CharPoly} has
$\mathcal{B}(\lambda_k)$ as a factor, vanishing at every point $(\lambda_k,w_k)$, for example with \eqref{w1PolyNLS} one
can easily compute:
\begin{multline*}
P(w_k,\lambda_k) = \frac{1}{\gamma_2^3(\lambda_k)}\Big(\big[\alpha_1(\lambda_k)+\alpha_2(\lambda_k)\big]\beta_1(\lambda_k) \gamma_2(\lambda_k) + \\ + \beta_3 (\lambda_k) \gamma_2^2(\lambda_k)
- \beta_1^2(\lambda_k)\gamma_3(\lambda_k) \Big)
\mathcal{B}(\lambda_k)\equiv 0.
\end{multline*}
Both the expressions for $\mathcal{A}$ give the same eigenvalue of the $\tens{L}$-matrix \eqref{muExpr}.

(ii) The assertion follows from the lemmas below.
Similar lemmas for a quad\-ra\-tic Poisson bracket can be found in \cite{Sklyanin92,Gekhtman}.

\begin{ConjVLemma}\label{L:ConjVarNLS}
If $\mathcal{B}$ and $\mathcal{A}$ satisfy the following identities with respect to
the first Lie-Poisson bra\-cket \eqref{PoissonBraNLS}
$$\{\mathcal{B}(u),\mathcal{B}(v)\}_{\textup{f}} = 0,\quad
\{\mathcal{A}(u),\mathcal{A}(v)\}_{\textup{f}} = 0, \quad
\{\mathcal{A}(u),\mathcal{B}(v)\}_{\textup{f}} = \frac{f(u,v)\mathcal{B}(u)-\mathcal{B}(v)}{u-v},$$
where $f$ is an arbitrary function such that $\lim_{v\to u} f(u,v)\,{=}\,1$,
then the variables $\{(\lambda_k,w_k)\,{:}$ $k\,{=}\,1,\,\dots,\,3\mathcal{N}\}$ defined by
$$\mathcal{B}(\lambda_k)=0,\qquad w_k=\mathcal{A}(\lambda_k)$$
are canonically conjugate with respect to $\{\cdot,\cdot\}_{\textup{f}}$:
$$\{\lambda_k,\lambda_l\}_{\textup{f}} = 0,\qquad
\{\lambda_k,w_l\}_{\textup{f}} = \delta_{kl},\qquad \{w_k,w_l\}_{\textup{f}} = 0. $$
\end{ConjVLemma}
\begin{proof}
The equations $\mathcal{B}(\lambda_k)\,{=}\,0$, $w_k \,{=}\, \mathcal{A}(\lambda_k)$ imply
\begin{align}\label{wlambdaDer}
&\frac{\partial \lambda_k}{\partial L_{a}^{(m)}} = - \lim_{u\to \lambda_k} \frac{1}{\mathcal{B}'(u)}
\frac{\partial \mathcal{B}(u)}{\partial L_{a}^{(m)}},\\
&\frac{\partial w_k}{\partial L_{a}^{(m)}}
= \lim_{u\to \lambda_k} \left(\frac{\partial \mathcal{A}(u)}{\partial L_{a}^{(m)}} +
\mathcal{A}'(u) \frac{\partial u}{\partial L_{a}^{(m)}} \right)
= \lim_{u\to \lambda_k} \left(\frac{\partial \mathcal{A}(u)}{\partial L_{a}^{(m)}} -
\frac{\mathcal{A}'(u)}{\mathcal{B}'(u)} \frac{\partial \mathcal{B}(u)}{\partial L_{a}^{(m)}} \right). \notag
\end{align}
Taking into account that
$$\sum_{m,n=0}^{\mathcal{N}-1}\sum_{a,b} \frac{\partial \mathcal{F}(u)}{\partial L_{a}^{(m)}} \frac{\partial \mathcal{G}(v)}{\partial L_{b}^{(n)}}
\{L_{a}^{(m)},L_{b}^{(n)}\}_{\text{f}} = \{\mathcal{F}(u),\mathcal{G}(v)\}_{\text{f}}$$
for any functions $\mathcal{F}$ and $\mathcal{G}$, one obtains the following
\begin{equation*}
\begin{split}
\{\lambda_k,\lambda_l\}_{\text{f}} &=
\frac{ \{\mathcal{B}(\lambda_k),\mathcal{B}(\lambda_l)\}_{\text{f}}}{\mathcal{B}'(\lambda_k)\mathcal{B}'(\lambda_l)} = 0,\\
\{w_k,\lambda_l\}_{\text{f}} &= \lim_{\substack{u\to \lambda_k \\ v\to \lambda_l}} \left( -\frac{1}{\mathcal{B}'(v)}\, \{\mathcal{A}(u),\mathcal{B}(v)\}_{\text{f}}
+ \frac{\mathcal{A}'(u)}{\mathcal{B}'(u) \mathcal{B}'(v)}\,
\{\mathcal{B}(u),\mathcal{B}(v)\}_{\text{f}}\right) =
\\ &=
-\frac{f(\lambda_k,\lambda_l)\mathcal{B}(\lambda_k) - \mathcal{B}(\lambda_l)}
{(\lambda_k-\lambda_l)\mathcal{B}'(\lambda_l)} = -\delta_{kl},\\
\{w_k,w_l\}_{\text{f}} &=
\lim_{\substack{u\to \lambda_k \\ v\to \lambda_l}}
\left(\frac{\mathcal{A}'(u)[f(v,u)\mathcal{B}(v)-\mathcal{B}(u)]}{\mathcal{B}'(u)(v-u)} -
\frac{\mathcal{A}'(v)[f(u,v)\mathcal{B}(u)-\mathcal{B}(v)]}{\mathcal{B}'(v)(u-v)}\right) =
\\ &=
\left(\frac{\mathcal{A}'(\lambda_k)}{\mathcal{B}'(\lambda_k)} -
\frac{\mathcal{A}'(\lambda_l)}{\mathcal{B}'(\lambda_l)}\right) \delta_{kl} = 0,
\end{split}
\end{equation*}
as required.
\end{proof}

\begin{ABLemma}\label{L:ABbracketNLS}
For $\mathcal{B}$ and $\mathcal{A}$ defined by \eqref{ABNLS} the following identities are true
with respect to the first Lie-Poisson bra\-cket \eqref{PoissonBraNLS}:
$$\{\mathcal{B}(u),\mathcal{B}(v)\}_{\textup{f}} = 0,\quad
\{\mathcal{A}(u),\mathcal{A}(v)\}_{\textup{f}} = 0, \quad
\{\mathcal{A}(u),\mathcal{B}(v)\}_{\textup{f}} = \frac{f(u,v)\mathcal{B}(u)-\mathcal{B}(v)}{u-v},$$
where $f(u,v)\,{=}\,\gamma_2^2(v)/\gamma_2^2(u)$ for \eqref{w1PolyNLS}, and
$f(u,v)\,{=}\,\gamma_3^2(v)/\gamma_3^2(u)$ for \eqref{w2PolyNLS}.
\end{ABLemma}
\begin{proof} It is realized by the direct computation.
From \eqref{PoissonBraNLS} written for polynomials as
$$\{L_{ij}(u), L_{kl}(v)\}_\text{f} = \frac{L_{kj}(u)-L_{kj}(v)}{u-v}\delta_{il} -
\frac{L_{il}(u)-L_{il}(v)}{u-v} \delta_{kj},$$
one obtains
\begin{align*}
&\{\gamma_{2}(u),\gamma_3(v)\} = \{\gamma_{2}(u),\Gamma_2(v)\} = \{\gamma_{3}(u),\Gamma_3(v)\} = 0,\\
&\{\gamma_{2}(u),\Gamma_3(v)\} = -\{\gamma_{3}(u),\Gamma_2(v)\} = \frac{1}{u-v} \begin{vmatrix}
\gamma_2(u) & \gamma_3(u) \\ \gamma_2(v) & \gamma_3(v) \end{vmatrix},\\
&\{\Gamma_{2}(u),\Gamma_3(v)\} = \frac{1}{u-v}\bigg( \begin{vmatrix}
\gamma_2(u) & \gamma_3(u) \\ \Gamma_2(v) & \Gamma_3(v) \end{vmatrix}
- \begin{vmatrix} \gamma_2(v) & \gamma_3(v) \\ \Gamma_2(u) & \Gamma_3(u) \end{vmatrix} \bigg),
\end{align*}
and then
\begin{gather*}
\begin{split}
&\{\gamma_{2}(u),\mathcal{B}(v)\} = \frac{\gamma_{2}(v)}{u-v} \begin{vmatrix}
\gamma_2(u) & \gamma_3(u) \\ \gamma_2(v) & \gamma_3(v) \end{vmatrix},\\
&\{\Gamma_{2}(u),\mathcal{B}(v)\} = \frac{1}{u-v} \bigg(\gamma_{2}(u) \mathcal{B}(v)-
 \gamma_{2}(v) \begin{vmatrix}
\gamma_2(v) & \gamma_3(v) \\ \Gamma_2(u) & \Gamma_3(u) \end{vmatrix} \bigg).
\end{split}\\
\Gamma_{2}(u)\{\gamma_{2}(u),\mathcal{B}(v)\} - \gamma_{2}(u)\{\Gamma_{2}(u),\mathcal{B}(v)\}
= \frac{1}{u-v} \big[\gamma^2_{2}(v) \mathcal{B}(u)-\gamma^2_{2}(u) \mathcal{B}(v)\big].
\end{gather*}
From Leibniz's rule for a Poisson bracket:
$$-\{\Gamma_{2}(u),\mathcal{B}(v)\} =
\{\mathcal{A}(u)\gamma_{2}(u),\mathcal{B}(v)\} =
\gamma_{2}(u)\{\mathcal{A}(u),\mathcal{B}(v)\} -
\frac{\Gamma_{2}(u)}{\gamma_{2}(u)} \{\gamma_{2}(u),\mathcal{B}(v)\},$$
where $\mathcal{A}(u)\,{=}\,{-}\Gamma_{2}(u)/\gamma_{2}(u)$, one easily gets
\begin{gather*}
\{\mathcal{A}(u),\mathcal{B}(v)\} =
\frac{1}{u-v}\Big(\frac{\gamma_2^2(v)}{\gamma^2_2(u)}\mathcal{B}(u)  - \mathcal{B}(v)\Big).
\end{gather*}
Other identities from the lemma statement are easily computed
in the similar way.
\end{proof}

\begin{remark}\label{r:ABlemma}
\nameABLemma\;\ref{L:ABbracketNLS} shows that \eqref{w1PolyNLS}, \eqref{w2PolyNLS}
give good expressions for $w(\lambda)$ because of canonical conjugation of $\mathcal{A}$ and $\mathcal{B}$
when the point $(\lambda,w)$ tends to $(\lambda_k,w_k)$ from the vicinity.
Explicit computation shows that solutions of \eqref{SpectCurveNLS} taken as the $\mathcal{A}$-function do not possess
this good property. The reason is that \eqref{SpectCurveNLS} is true only at the points
$\{(\lambda_k,w_k)\}$ but not in their vicinities.
In other words, the $\mathcal{A}$-function defined by \eqref{ABNLS} conserves canonical conjugation with
the polynomial $\mathcal{B}$ when the dynamic variables evolve, but the solutions of \eqref{SpectCurveNLS}
do not.
\end{remark}

(iii) The Liouville 1-form on $\mathcal{O}_{\text{f}}$ is
implied by (\ref{PoissonBraNLSConj}):
\begin{equation*}
\Omega_{\text{f}}=\sum\limits_{k} w_{k}\,d\lambda_{k}.
\end{equation*}
Fixing values of the Hamiltonians $h_0$, $h_1$, \ldots, $h_{\mathcal{N}-1}$, $f_0$, $f_1$, \ldots, $f_{2\mathcal{N}-1}$
we obtain a Liouville torus.
On the torus every variable $w_k$ becomes an algebraic function of the corresponding conjugate variable $\lambda_k$
due to (\ref{HyperCurveNLS}),
and the form $\Omega_{\text{f}}$ becomes a sum of
meromorphic differentials on the Riemann surface $P(w,\lambda)=0$.

This completes the proof of \nameSoV\;\ref{T:SoVNLS}.
\end{proof}

In order to obtain the required $3\mathcal{N}$ points $\{(\lambda_k,w_k)\}$ we need the polynomial~$\mathcal{B}$
of degree $3\mathcal{N}$, this is provided by the maximal degrees of the polynomials $\Gamma_2$ and $\gamma_3$,
or $\Gamma_3$ and $\gamma_2$. If the leading coefficient $\tens{L}^{(\mathcal{N})}$ of the Lax matrix~$\tens{L}$
defined by \eqref{muExpr} does not provide the maximal degrees,
one can apply a proper similarity transformation to $\tens{L}$.

Inversely, given a set of pairs $\{(\lambda_k, w_k)\,{:}\,k\,{=}\,1,\,\dots,\, 3\mathcal{N}\}$ one computes
the dynamic variables $\{\gamma_m^{(1)}$, $\beta_m^{(1)}$, $\alpha_{m}^{(1)}$, $\gamma_m^{(3)}$, $\gamma_m^{(2)}$,
$\alpha_{m}^{(2)}\,{:}$ $m\,{=}\,0$, $1,\,\dots$, $\mathcal{N}\}$ such that the equations
(\ref{newvarNLS}) are satisfied. Thus, one defines a homomorphism
\begin{equation}\label{homomorfizmNLS}
  \Complex^{6\mathcal{N}} \to \mathcal{O}_\text{f}
\end{equation}
that maps $\{(\lambda_k, w_k)\}$ to a point of $\mathcal{O}_\text{f}$.
When all the Hamiltonians are fixed the homomorphism \eqref{homomorfizmNLS} turns into the
map from the symmetrized product of $3\mathcal{N}$ Riemann surfaces $\mathcal{R}$
defined by \eqref{HyperCurveNLS} to the Liouville torus:
\begin{equation*}
\Sym\{\mathcal{R}\times \mathcal{R}\times \cdots \times
\mathcal{R}\} \mapsto T^{3\mathcal{N}}.
\end{equation*}

\begin{remark}
One can observe a mnemonic rule: the expressions \eqref{ABNLS} are easily obtained
from \eqref{InvarF} written in the matrix form
$$
\begin{pmatrix} I_2(\lambda) \\ I_3(\lambda) \end{pmatrix}  =
\begin{pmatrix} \gamma_2(\lambda) & \gamma_3(\lambda) \\
\Gamma_2(\lambda) & \Gamma_3(\lambda) \end{pmatrix}
\begin{pmatrix} \beta_2(\lambda) \\ \beta_3(\lambda) \end{pmatrix} +
\begin{pmatrix}  \alpha_2^2(\lambda) - A(\lambda)  \\ - \alpha_2(\lambda) A(\lambda)  \end{pmatrix}.
$$
The consistent equation~\eqref{ConsistentEqNLS} provides singularity of the matrix $(\begin{smallmatrix}
\gamma_2 & \gamma_3 \\ \Gamma_2 &\Gamma_3 \end{smallmatrix})$ at every point $(\lambda_k,w_k)$,
thus the spectral curve \eqref{SpectrCurve}
is reduced to the form \eqref{SpectCurveNLS}.
\end{remark}

\section{Connection to the well-known results}
Considering separation of variables for integrable systems on orbits in loop algebras
we refer the papers \cite{Sklyanin92, Adams, AdamsCO, Scott, Sklyanin95, Gekhtman}.

We start from Sklyanin's paper \cite{Sklyanin92} where the representation \eqref{newvarNLS} for
variables of separation was declared firstly as a conjecture,
and proven for the classical SL(3) magnetic chain. In \cite{Gekhtman} this assertion was extended
to the classical SL($n$) magnetic chain. The system was considered
in the phase space with a quadratic Poisson bracket, but separation of variables
was realized by the expressions similar to \eqref{ABNLS}.
This is presumably true for any integrable system on coadjoint orbits of the $\mathfrak{sl}(3)$ loop algebra.
As shown below the variables of separation on orbits of the second type are defined by the same expressions.

For further explanation we introduce the matrix  $\tens{N}(\lambda,w)\,{\equiv}\,\tens{L}(\lambda)\,{-}\,w\Ibb$
with the $\tens{L}$-matrix \eqref{muExpr}, and denote by $\widetilde{\tens{N}}$ its adjoint matrix
whose entries $\widetilde{N}_{ij}$ are cofactors of $N_{ji}$.
One can easily see that \eqref{Minors23} is equivalent to
\begin{gather*}
\widetilde{N}_{31} (\lambda_k,w_k)  \equiv \begin{vmatrix}
\gamma_1(\lambda_k) & \alpha_2(\lambda_k) - \alpha_1(\lambda_k) - w_k \\
\gamma_3(\lambda_k) & \gamma_2(\lambda_k)
\end{vmatrix} = 0,\\
\widetilde{N}_{32} (\lambda_k,w_k)  \equiv - \begin{vmatrix}
\alpha_1(\lambda_k) - w_k & \beta_1(\lambda_k) \\ \gamma_3(\lambda_k) & \gamma_2(\lambda_k)
\end{vmatrix} = 0,
\end{gather*}
which gives expressions for the $\mathcal{A}$-function.
At the same time elimination of $\{w_k\}$ leads to the polynomial $\mathcal{B}$,
giving the consistent condition \eqref{ConsistentEqNLS}.
In \cite{Sklyanin95} Sklyanin presented this result as is;
however it naturally follows from the orbit method.
Considering the manifold $\mathcal{M}_0$ as a foliation into generic orbits
we obtain expressions for $\mathcal{A}$ and $\mathcal{B}$ from
the relations between dynamic and spectral variables.

In addition we compute the eigenvector of $\tens{L}(\lambda_k)$
corresponding to the eigenvalue $w_k$ defined by \eqref{w1PolyNLS} or \eqref{w2PolyNLS}.
Using  \eqref{Minors23} and supposing $\lambda_k$ does not coincide with any root
of $\gamma_2$ and $\gamma_3$, by Gauss' method one reduces the matrix $\tens{N}(\lambda_k,w_k)$ to the form
$$
\begin{pmatrix}
-\frac{1}{\gamma_2(\lambda_k)} \widetilde{N}_{32}(\lambda_k) &0
&\frac{1}{\gamma_2(\lambda_k)}\widetilde{N}_{12}(\lambda_k) \\
0&-\frac{1}{\gamma_3(\lambda_k)} \widetilde{N}_{31}(\lambda_k)
&\frac{1}{\gamma_3(\lambda_k)}\widetilde{N}_{21}(\lambda_k) \\
\gamma_3(\lambda_k) & \gamma_2(\lambda_k) & -\alpha_2(\lambda_k) - w_k
\end{pmatrix}$$
with the vanishing left upper $2\,{\times}\,2$ block, and the nonvanishing last column. The corresponding
eigenvector has the form $\vec{\Omega}^{\text{T}} \,{=}\, \big(\Omega_1,\, \Omega_2,\, 0\big)$ such that
\begin{equation}\label{NormCond}
\gamma_3(\lambda_k) \Omega_1 \,{+}\, \gamma_2(\lambda_k) \Omega_2 \,{=}\, 0,
\end{equation}
then
\begin{equation}\label{EigenVect}
\vec{\Omega}^{\text{T}} \,{=}\,\Omega_1 \big(1,\,-\gamma_3(\lambda_k)/\gamma_2(\lambda_k),0\big)\quad
\text{or} \quad \Omega_2 \big(-\gamma_2(\lambda_k)/\gamma_3(\lambda_k),1,\,0\big).
\end{equation}
One can also observe that the polynomial $\mathcal{B}$ has the form of \eqref{NormCond},
and the vector $$ \begin{pmatrix}
\Gamma_2(\lambda_k) \\ -\Gamma_3(\lambda_k) \\ 0 \end{pmatrix} =
w_k \begin{pmatrix} -\gamma_2(\lambda_k) \\ \gamma_3(\lambda_k)\\ 0 \end{pmatrix} $$ at every root of $\mathcal{B}$
serves as an eigenvector for $\tens{L}(\lambda_k)$, here we use the relations \eqref{Minors23}.
If $\mathcal{B}$ has the maximal degree $3\mathcal{N}$ then
there exist $3\mathcal{N}$ values of $\{\lambda_k\}$ satisfying the consistent condition \eqref{ConsistentEqNLS},
and the corresponding values of $\{w_k\}$. Note that we take only one eigenvalue $w_k$ for every~$\lambda_k$.
Recall that coefficients of $\Gamma_2$, $\Gamma_3$, $\gamma_2$, $\gamma_3$ are not constant but serve as dynamic variables of the system in question. Their evolution implies an evolution of the the spectral variables $(\lambda,\,w)$, and
it is convenient to shift the focus from the dynamic variables onto the spectral ones because
the latter are canonically conjugate. The relations \eqref{Minors23} fix
an unambiguous connection between dynamic and spectral variables. It means we take a sufficient number of points $(\lambda_k,\,w_k)$
such that the spectral curve has a simple form: among $6\mathcal{N}$ branch points $4\mathcal{N}$
pairwise coincide. Conserving this property, the curve evolves together with the dynamic variables.

Now we obtain the matrix $\tens{K}$, introduced in \cite{Sklyanin92}, which realizes
the similarity transformation reducing the $\tens{L}$-matrix
to a block-triangular form. A certain entry of this block-triangular form is used
to obtain the polynomial $\mathcal{B}$. In \cite{Scott} this idea of
constructing the polynomial $\mathcal{B}$ was developed for the SL($n$) case.
Using \eqref{EigenVect} one immediately writes the transformation matrix $\tens{K}$
and the corresponding transformation of~$\tens{L}$:
\begin{gather*}
\tens{K}_1 = \small
\begin{pmatrix} 1&0&0\\ -\gamma_3(\lambda_k)/\gamma_2(\lambda_k)&1&0
\\ 0&0&1 \end{pmatrix},\quad
\tens{K}_1 \tens{L} \tens{K}_1^{-1} =
\begin{pmatrix} \alpha_1 - \frac{\beta_1 \gamma_3}{\gamma_2}&\beta_1&\beta_3\\
\mathcal{B}/\gamma_2^2&\alpha_2-\alpha_1+\frac{\beta_1 \gamma_3}{\gamma_2}&
\beta_2+\frac{\beta_3 \gamma_3}{\gamma_2} \\ 0&\gamma_2&-\alpha_2 \end{pmatrix}
\end{gather*}
or
\begin{gather*}
\tens{K}_2 = \small \begin{pmatrix} 1&-\gamma_2(\lambda_k)/\gamma_3(\lambda_k)&0\\
0&1&0 \\ 0&0&1 \end{pmatrix},\quad
\tens{K}_2 \tens{L} \tens{K}_2^{-1} = \begin{pmatrix}
\alpha_1 + \frac{\gamma_1 \gamma_2}{\gamma_3}&-\mathcal{B}/\gamma_3^2&
\beta_3+\frac{\beta_2\gamma_2}{\gamma_3}\\
\gamma_1&\alpha_2-\alpha_1-\frac{\gamma_1 \gamma_2}{\gamma_3}& \beta_2\\
\gamma_3&0&-\alpha_2 \end{pmatrix}.
\end{gather*}
It is easy to extract the polynomial $\mathcal{B}$ whose vanishing detects the eigenvalue given by
\eqref{w1PolyNLS} or \eqref{w2PolyNLS} respectively.
These transformations of $\tens{L}$ show that \emph{all points $\{(\lambda_k,w_k)\}$ belong to
the same sheet of the trigonal curve \eqref{SpectrCurve}}.

In \cite{Adams} and more detailed in \cite{AdamsCO} Adams, Harnad and Hurtubise showed that variables of separation, called spectral Darboux coordinates, are zeros of $\widetilde{\tens{N}}(\lambda,w)\bm{v}_0$ with an arbitrary vector $\bm{v}_0$ usually chosen as $(1,\,0,\,\dots,\,0)^\text{T}$. The variables form the divisor of a section of the eigenvector line bundle over
the invariant spectral curve corresponding to a system. Applying this idea to the above system on $\mathcal{O}_{\text{f}}$
we get the equations
$$\widetilde{\tens{N}}(\lambda,w) \small \begin{pmatrix} 1 \\ 0 \\ 0 \end{pmatrix} =
\begin{pmatrix} \widetilde{N}_{11}(\lambda,w) \\ \widetilde{N}_{21}(\lambda,w) \\
\widetilde{N}_{31}(\lambda,w) \end{pmatrix} = 0,$$
which take place if one eliminates the set $\{\beta_{1}^{(m)}, \beta_{3}^{(m)}{:}\,m\,{=}\,0,\,1$,
\ldots, $\mathcal{N}\}$ of the dynamic variables.
The equations $\widetilde{N}_{21}(\lambda,w)\,{=}\,0$, $\widetilde{N}_{31}(\lambda,w)\,{=}\,0$
give expressions for $\mathcal{A}$ like \eqref{w1PolyNLS}, \eqref{w2PolyNLS}. The first
equation is a simplification of a spectral curve equation like \eqref{SpectCurveNLS}, true only for the set $\{(\lambda_k,w_k)\}$
satisfying both the other two equations. Note that the first equation $\widetilde{N}_{11}(\lambda,w)\,{=}\,0$
can not be used to define $w(\lambda)$, though this is a characteristic equation.
The reason is the absence of desirable property of canonical conjugation (see Remark\;\ref{r:ABlemma}).
Only expressions \eqref{w1PolyNLS}, \eqref{w2PolyNLS} meet this requirement as \nameABLemma\;1 shows.

To complete the comparison with the results from \cite{AdamsCO} we note that the proposed procedure of
separation of variables gives the equations \eqref{Minors23}, \eqref{SpectCurveNLS} equivalent to
$$\widetilde{\tens{N}}^\text{T}(\lambda,w)\small \begin{pmatrix} 0 \\ 0 \\ 1 \end{pmatrix} = 0.$$
The reader can see that such section of the dual eigenvector line bundle over the spectral curve
is also acceptable.

Solvability of the equations for the spectral coordinates $\{(\lambda_k,w_k)\}$ is a delicate question
related to the leading coefficient $\tens{L}^{(\mathcal{N})}$ of the $\tens{L}$-matrix. In \cite{Adams,AdamsCO} it
was formulated in terms of the vector $\bm{v}_0$: if $\bm{v}_0$ is an eigenvector of $\tens{L}^{(\mathcal{N})}$
the equations $\widetilde{\tens{N}}(\lambda,w)\bm{v}_0\,{=}\,0$ give only $3\mathcal{N}\,{-}\,2$ spectral points $\{(\lambda_k,w_k)\}$,
this number coincides with the genus of the spectral curve. The two missing points lie over $\lambda\,{=}\,\infty$.
Adams, Harnad, Hurtubise gave a rule how to construct the complete set of variables of separation
in this special case. In the proposed procedure of separation of variables this question
arises if the degree of $\mathcal{B}$ is less then $3\mathcal{N}$, then a proper similarity transformation
of the $\tens{L}$-matrix solves the problem.

We see that all ideas of constructing variables of separation receive simple and obvious explanations
by means of the orbit method. It allows to obtain the relations producing variables of separation
in a natural way by restriction to an orbit and changing variables
from dynamic to spectral.

\section{Separation of variables on $\mathcal{O}_{\text{s}}$}
The orbit $\mathcal{O}_{\text{s}}$ equipped with the second Lie-Poisson bracket \eqref{LiePoissonBraHM}
has the Poisson structure:
\begin{gather}\label{PoissonBraHM}
\{L_{ij}^{(m)}, L_{kl}^{(n)}\}_{\text{s}} = L_{kj}^{(m+n+1-\mathcal{N})}\delta_{il}
- L_{il}^{(m+n+1-\mathcal{N})} \delta_{kj},
\end{gather}
or in terms of the $\tens{r}$-matrix \eqref{Rmatrix}
\begin{gather*}
\{\tens{L}_1(u)\overset{\otimes}{,}\, \tens{L}_{2}(v)\}_{\text{s}} = -
[\tens{r}_{12}(u-v),v^{\mathcal{N}}\tens{L}_1(u)+u^{\mathcal{N}}\tens{L}_{2}(v)].
\end{gather*}

We parameterize the orbit $\mathcal{O}_{\text{s}}$ by the same dynamic
variables $\{\gamma_1^{(m)}$, $\gamma_2^{(m)}$, $\gamma_3^{(m)}$,
$\beta_1^{(m)}$, $\alpha_{1}^{(m)}$,  $\alpha_{2}^{(m)}{:}$ $m\,{=}\,0$, $1$, \ldots, $\mathcal{N}\,{-}\,1\}$,
namely: we again eliminate the set $\{\beta_2^{(m)}, \beta_3^{(m)}\}$. Due
to linearity of the orbit equations \eqref{OrbEqHM} in the eliminated variables
we write them in the matrix form
\begin{gather}\label{betaSystHM}
\vec{c}_{\text{s}} = \tens{S}^- \vec{\beta} + \vec{\eta}^{-}_\text{s},
\end{gather}
where
\begin{gather*}
\tens{S}^{-}=\small \begin{bmatrix}
   \tens{S}_{0} & 0 & \dots & 0 & 0 \\
   \tens{S}_{1} & \tens{S}_{0} & \dots & 0 & 0 \\
   \vdots & \vdots& \ddots& \vdots& \vdots \\
   \tens{S}_{\mathcal{N}-1} & \tens{S}_{\mathcal{N}-2} & \dots & \tens{S}_{0} & 0 \\
   0 & 0 & \dots & 0 & \tens{F}_{\mathcal{N}}
   \end{bmatrix},\quad
\vec{\beta} = \begin{bmatrix} \vec{\beta}_{0} \\ \vec{\beta}_{1} \\ \vdots \\
   \vec{\beta}_{\mathcal{N}-1} \\ \vec{\beta}_{\mathcal{N}}  \end{bmatrix}, \quad
\vec{c}_{\text{s}} = \begin{bmatrix} \vec{c}_{0}^{\text{s}} \\ \vec{c}_{1}^{\text{s}} \\ \vdots \\
   \vec{c}_{\mathcal{N}-1}^{\text{s}} \\ \vec{c}_{2\mathcal{N}}^{\text{f}} \end{bmatrix}, \quad
\vec{\eta}^{-}_\text{s} = \begin{bmatrix} \vec{\eta}_{0}^{\text{s}} \\ \vec{\eta}_{1}^{\text{s}} \\ \vdots \\
   \vec{\eta}_{\mathcal{N}-1}^{\text{s}} \\ \vec{\eta}_{2\mathcal{N}}^{\text{f}} \end{bmatrix},\\
\tens{S}_{j} = \small \begin{bmatrix} \gamma_2^{(j)} &  \gamma_3^{(j)} \\ \Gamma_2^{(j)}  & \Gamma_3^{(j)} \end{bmatrix},\quad
\vec{\beta}^{(j)} = \begin{bmatrix} \beta_2^{(j)} \\ \beta_3^{(j)} \end{bmatrix}, \quad
\vec{c}_{j}^{\text{s}} = \begin{bmatrix} c_{j} \\ d_{j} \end{bmatrix}, \quad
\vec{\eta}_{j}^{\text{s}} = \begin{bmatrix} \eta_{j} \\ H_{j}  \end{bmatrix}.
\end{gather*}
Supposing $\tens{S}_{0}$ is nonsingular, we eliminate the variables $\vec{\beta}$
\begin{gather*}
\vec{\beta} = (\tens{S}^{-})^{-1} (\vec{c}_{\text{s}} - \vec{\eta}^{-}_\text{s}), \qquad
\text{or} \\ \small
\begin{bmatrix} \vec{\beta}_{0} \\ \vec{\beta}_{1} \\ \vdots \\
   \vec{\beta}_{\mathcal{N}-1} \\ \vec{\beta}_{\mathcal{N}}  \end{bmatrix} =
   \begin{bmatrix}
   \tens{S}_{0}^{-1} & 0 & \dots & 0 & 0 \\
   \widetilde{\tens{S}}_{1} & \tens{S}_{0}^{-1} & \dots & 0 & 0 \\
   \vdots & \vdots& \ddots& \vdots& \vdots \\
   \widetilde{\tens{S}}_{\mathcal{N}-1} &
   \widetilde{\tens{S}}_{\mathcal{N}-2} & \dots & \tens{S}_{0}^{-1} & 0 \\
   0 & 0 & \dots & 0 & \tens{F}_{\mathcal{N}}^{-1}
   \end{bmatrix}
   \begin{bmatrix}
   \vec{c}_{0}^{\text{s}} - \vec{\eta}_{0}^{\text{s}} \\
   \vec{c}_{1}^{\text{s}} - \vec{\eta}_{1}^{\text{s}} \\ \vdots \\
   \vec{c}_{\mathcal{N}-1}^{\text{s}} - \vec{\eta}_{\mathcal{N}-1}^{\text{s}}  \\
   \vec{c}_{2\mathcal{N}}^{\text{f}} - \vec{\eta}_{2\mathcal{N}}^{\text{f}}
   \end{bmatrix},\\
   \widetilde{\tens{S}}_{n} = \tens{S}_{0}^{-1} \sum_{k=1}^n
   \big(-\tens{S}_{n+1-k} \tens{S}_{0}^{-1}\big)^k, \qquad n=1,\,\dots,\,\mathcal{N}-1.
\end{gather*}
Then substitute $\vec{\beta}$ into  the Hamiltonians $h_{\mathcal{N}}, h_{\mathcal{N}+1}$, $\ldots$,
$h_{2\mathcal{N}-1}$, $f_{\mathcal{N}}$, $\ldots$,  $f_{3\mathcal{N}-1}$
\begin{equation}\label{HamiltHM}
   \vec{h}_{\text{s}} = \tens{S}^{+} \vec{\beta} +  \vec{\eta}^{+}_\text{s} =
   \tens{S}^{+} (\tens{S}^{-})^{-1} \vec{c}_{\text{s}} + \vec{\eta}^{+}_\text{s}
   - \tens{S}^{+} (\tens{S}^{-})^{-1} \vec{\eta}^{-}_\text{s},
\end{equation}
where
\begin{gather*}
   \small \tens{S}^{+} = \begin{bmatrix}
   \tens{S}_{\mathcal{N}} & \tens{S}_{\mathcal{N}-1} & \dots & \tens{S}_{1} & \tens{S}_{0} \\
   \tens{g}_{\mathcal{N}+1}^0 & \tens{S}_{\mathcal{N}} & \dots & \tens{S}_{2} & \tens{S}_{1} \\
   \vdots& \vdots& \ddots& \vdots& \vdots \\
   \tens{g}_{2\mathcal{N}-1}^0 & \tens{g}_{2\mathcal{N}-2}^0 & \dots & \tens{S}_{\mathcal{N}} &
   \tens{S}_{\mathcal{N}-1} \\
   \tens{g}_{2\mathcal{N}} &  \tens{g}_{2\mathcal{N}-1} & \dots & \tens{g}_{\mathcal{N}+1}
   & \tens{g}_{\mathcal{N}} \\
   0 & \tens{g}_{2\mathcal{N}} & \dots & \tens{g}_{\mathcal{N}+2} & \tens{g}_{\mathcal{N}+1} \\
   \vdots& \vdots& \ddots& \vdots &\vdots \\
   0 & 0 & \dots & \tens{g}_{2\mathcal{N}} & \tens{g}_{2\mathcal{N}-1} \\
   \end{bmatrix}\!,\ \
\vec{h}_{\text{s}} = \begin{bmatrix}
   \vec{h}_{\mathcal{N}}^{\text{s}} \\ \vec{h}_{\mathcal{N}+1}^{\text{s}} \\  \vdots \\
   \vec{h}_{2\mathcal{N}-1}^{\text{s}} \\
   f_{2\mathcal{N}} \\ f_{2\mathcal{N}+1} \\ \vdots \\  f_{3\mathcal{N}-1} \end{bmatrix}\!,  \ \
\vec{\eta}^{+}_\text{s} = \begin{bmatrix}
   \vec{\eta}_{\mathcal{N}}^{\text{s}} \\ \vec{\eta}_{\mathcal{N}+1}^{\text{s}} \\  \vdots \\
   \vec{\eta}_{2\mathcal{N}-1}^{\text{s}} \\
   H_{2\mathcal{N}} \\ H_{2\mathcal{N}+1} \\ \vdots \\ H_{3\mathcal{N}-1} \\  \end{bmatrix}\!,\ \
\vec{h}_j^{\text{s}} = \begin{bmatrix} h_{j} \\ f_{j} \end{bmatrix}.
\end{gather*}
Note that the expressions (\ref{HamiltHM}) are linear in
$\{c_{\nu}$, $d_{\nu}\,{:}$ $\nu\,{=}\,0, \ldots, \mathcal{N}\,{-}\,1\}$.

Now consider the spectral curve restricted to the orbit $\mathcal{O}_\text{s}$ defined by \eqref{OrbEqHM}.
We write the following set of equations for $3\mathcal{N}$ points $\{(\lambda_k,\,
w_k)\}$ on the orbit
\begin{multline}\label{HyperCurveHM}
  w_k^3 = w_k\Big(c_{0}+c_1 \lambda_k + \cdots +
  c_{\mathcal{N}-1} \lambda_k^{\mathcal{N}-1} + h_{\mathcal{N}} \lambda_k^{\mathcal{N}} + \cdots + h_{2\mathcal{N}-1} \lambda_k^{2\mathcal{N}-1}
  + c_{2\mathcal{N}} \lambda^{2\mathcal{N}}\Big)
  + \\ +\Big(d_{0} + d_1 \lambda_k + \cdots + d_{\mathcal{N}-1}\lambda_k^{\mathcal{N}-1} +
  f_{\mathcal{N}} \lambda_k^{\mathcal{N}} + \cdots + f_{3\mathcal{N}-1} \lambda_k^{3\mathcal{N}-1}
  + d_{3\mathcal{N}} \lambda_k^{3\mathcal{N}}\Big)
\end{multline}
or in the matrix form
\begin{gather*}
\tens{W}^-_\text{s} \vec{c}_\text{s} +  \tens{W}^+_\text{s} \vec{h}_\text{s} = \vec{w}^{\text{cubed}}, \\
\small
\tens{W}^-_\text{s} = \begin{bmatrix}
\tens{W}_1 & \lambda_1 \tens{W}_1 & \dots & \lambda_1^{\mathcal{N}-1} \tens{W}_1 & \lambda_1^{2\mathcal{N}} \tens{W}_1 \\
\tens{W}_2 & \lambda_2 \tens{W}_2 & \dots & \lambda_2^{\mathcal{N}-1} \tens{W}_2 & \lambda_2^{2\mathcal{N}} \tens{W}_2 \\
\vdots & \vdots & \dots & \vdots & \vdots \\
\tens{W}_{3\mathcal{N}} & \lambda_{3\mathcal{N}} \tens{W}_{3\mathcal{N}} & \dots &
\lambda_{3\mathcal{N}}^{\mathcal{N}-1} \tens{W}_{3\mathcal{N}} &
\lambda_{3\mathcal{N}}^{\mathcal{N}} \tens{W}_{3\mathcal{N}} \end{bmatrix},
\quad \tens{W}_k \,{=}\, \begin{bmatrix} w_k & 1 \end{bmatrix},\\
\small \tens{W}^+_\text{s} = \begin{bmatrix}
\lambda_1^{\mathcal{N}} \tens{W}_1 & \lambda_1^{\mathcal{N}+1} \tens{W}_1 & \dots & \lambda_1^{2\mathcal{N}-1} \tens{W}_1
& \lambda_1^{2\mathcal{N}} & \lambda_1^{2\mathcal{N}+1} & \dots & \lambda_1^{3\mathcal{N}-1} \\
\lambda_2^{\mathcal{N}} \tens{W}_2 & \lambda_2^{\mathcal{N}+1} \tens{W}_2 & \dots & \lambda_2^{2\mathcal{N}-1} \tens{W}_2
& \lambda_2^{2\mathcal{N}} & \lambda_2^{2\mathcal{N}+1} & \dots & \lambda_2^{3\mathcal{N}-1} \\
\vdots & \vdots & \dots & \vdots & \vdots & \vdots & \dots & \vdots \\
\lambda_{3\mathcal{N}}^{\mathcal{N}} \tens{W}_{3\mathcal{N}} & \lambda_{3\mathcal{N}}^{\mathcal{N}+1} \tens{W}_{3\mathcal{N}}
& \dots & \lambda_{3\mathcal{N}}^{2\mathcal{N}-1} \tens{W}_{3\mathcal{N}} & \lambda_{3\mathcal{N}}^{2\mathcal{N}}
& \lambda_{3\mathcal{N}}^{2\mathcal{N}+1} & \dots & \lambda_{3\mathcal{N}}^{3\mathcal{N}-1} \end{bmatrix}.
\end{gather*}
Suppose all pairs $\{(\lambda_k,w_k)\}$ are distinct points and
the matrix $\tens{W}^+_{\text{s}}$ is nonsingular, then the Hamiltonians can be computed by
the formula
\begin{equation}\label{hSystHM}
 \vec{h}_{\text{s}}  =  - (\tens{W}^+_{\text{s}})^{-1} \tens{W}^-_{\text{s}} \vec{c}_{\text{s}} + (\tens{W}^+_{\text{s}})^{-1}\vec{w}^{\text{cubed}}.
\end{equation}
On an orbit $\mathcal{O}_\text{s}$ of the second type the formulas (\ref{HamiltHM}) and (\ref{hSystHM})
define the same set of functions, and both
of them are linear in $\{c_{\nu},\,d_{\nu}\,{:}\,
\nu\,{=}\,0,\, \dots,\, \mathcal{N}\}$. As $\{c_{\nu},\,d_{\nu}\}$ are
independent parameters one can equate the corresponding
terms, that is
\begin{gather}
\tens{S}^{+} (\tens{S}^{-})^{-1} = - (\tens{W}^+_{\text{s}})^{-1} \tens{W}^-_{\text{s}},\qquad
 \vec{\eta}^{+}_{\text{s}} - \tens{S}^{+} (\tens{S}^{-})^{-1} \vec{\eta}^{-}_{\text{s}} =
 (\tens{W}^+_{\text{s}})^{-1} \vec{w}^{\text{cubed}}
 \quad \Rightarrow \nonumber\\
 \tens{W}^+_{\text{s}} \tens{S}^{+}  + \tens{W}^-_{\text{s}} \tens{S}^{-} = 0,\qquad
 \tens{W}^+_{\text{s}} \vec{\eta}^{+}_{\text{s}} + \tens{W}^-_{\text{s}}
 \vec{\eta}^{-}_{\text{s}} = \vec{w}^{\text{cubed}}. \label{RootEqHM}
\end{gather}
The matrix equations \eqref{RootEqHM} give the equations \eqref{Minors23},
connecting the dynamic variables $\{\vec{\gamma}_2$, $\vec{\Gamma}_2$,
$\vec{\gamma}_3$, $\vec{\Gamma}_3\}$ with the spectral variables $\{\vec{\lambda},\,\vec{w}\}$,
and the simplification \eqref{SpectCurveNLS} of the spectral curve equation
at the points $\{(\lambda_k,w_k)\}$.
Evidently, we obtain the consistent equation \eqref{ConsistentEqNLS},
defining the polynomial~$\mathcal{B}$.

\begin{SoV}\label{T:SoVHM}
Suppose the orbit $\mathcal{O}_{\text{s}}$ is parameterized by the variables
 $\{\gamma_1^{(m)}$, $\beta_1^{(m)}$, $\alpha_1^{(m)}$, $\gamma_3^{(m)}$, $\gamma_2^{(m)}$,
 $\alpha_2^{(m)}{:}$ $m\,{=}\,0,\,\dots,\,\mathcal{N}\,{-}\,1\}$ as above.
Then the new variables $\{(\lambda_k, w_k)\,{:}$ $k\,{=}\,1,\,\dots,\,3\mathcal{N}\}$
defined by the formulas
\begin{equation}\label{newvarHM}
  \mathcal{B}(\lambda_k)=0,\qquad
  w_k = \mathcal{A}(\lambda_k),
\end{equation}
where $\mathcal{B}$ is the polynomial of degree $3\mathcal{N}$ and $\mathcal{A}$
is the algebraic function given by the expressions \eqref{ABNLS},
have the following properties:
\begin{enumerate}
\item[\textup{(i)}]  a pair $(\lambda_k,w_k)$ is a root of the characteristic
polynomial \eqref{CharPoly}.
\item[\textup{(ii)}] a pair
$(\lambda_k, w_k)$ is quasi-canonically conjugate with respect to
the second Lie-Poisson bracket \eqref{LiePoissonBraHM}:
\begin{equation}\label{PoissonBraHMConj}
  \{\lambda_k,\lambda_l\}_{\textup{s}} =0 \qquad
  \{\lambda_k, w_l\}_{\textup{s}} = -\lambda_k^{\mathcal{N}}\delta_{kl}, \qquad
  \{w_k,w_l\}_{\textup{s}}=0;
\end{equation}
\item[\textup{(iii)}] the corresponding  Liouville 1-form is
\begin{gather*}
\Omega_{\textup{s}}= - \sum\limits_{k} \lambda_k^{-\mathcal{N}} w_{k}\,d\lambda_{k}.
\end{gather*}
\end{enumerate}
\end{SoV}

\begin{proof}
(i) The proof repeats one for \nameSoV\,\ref{T:SoVNLS}.

(ii) The assertion follows from the lemmas below.
\begin{ConjVLemma}\label{L:ConjVarHM}
If $\mathcal{B}$ and $\mathcal{A}$ satisfy the following identities
with respect to the second Lie-Poisson bracket \eqref{PoissonBraHM}
\begin{gather*}
\{\mathcal{B}(u),\mathcal{B}(v)\}_{\textup{s}} = 0,\quad
\{\mathcal{A}(u),\mathcal{A}(v)\}_{\textup{s}} = 0, \\
\{\mathcal{A}(u),\mathcal{B}(v)\}_{\textup{s}} = -
\frac{f(u,v) v^{\mathcal{N}}\mathcal{B}(u)-u^{\mathcal{N}}\mathcal{B}(v)}{u-v},
\end{gather*}
where $f$ is an arbitrary function such that $\lim_{v\to u} f(u,v)\,{=}\,1$,
then the variables $\{(\lambda_k,w_k)\}$ defined by
$$\mathcal{B}(\lambda_k)=0,\qquad w_k=\mathcal{A}(\lambda_k)$$
are quasi-canonically conjugate with respect to $\{\cdot,\cdot\}_{\textup{s}}$:
$$\{\lambda_k,\lambda_l\}_{\textup{s}} = 0,\qquad \{\lambda_k,w_l\}_{\textup{s}} =
-\lambda_k^{\mathcal{N}}\delta_{kl},\qquad \{w_k,w_l\}_{\textup{s}} = 0.$$
\end{ConjVLemma}
\begin{proof} It is similar to the proof of \nameConjVLemma\;\ref{L:ConjVarNLS}.
Using \eqref{wlambdaDer} one can easily compute
\begin{equation*}
\begin{split}
&\{\lambda_k,\lambda_l\}_{\text{s}} =
\frac{\{\mathcal{B}(\lambda_k),\mathcal{B}(\lambda_l)\}_{\text{s}}}{\mathcal{B}'(\lambda_k)\mathcal{B}'(\lambda_l)} = 0,\\
&\{w_k,\lambda_l\}_{\text{s}} =
\lim_{\substack{u\to \lambda_k \\ v\to \lambda_l}} \left(- \frac{1}{\mathcal{B}'(v)}\,
\{\mathcal{A}(u),\mathcal{B}(v)\}_{\text{s}}
+ \frac{\mathcal{A}'(u)}{\mathcal{B}'(u) \mathcal{B}'(v)}\, \{\mathcal{B}(u),\mathcal{B}(v)\}_{\text{s}}\right) = \\
&\phantom{\{\lambda_k,\lambda_l\}_{\text{s}}} = \frac{f(\lambda_k,\lambda_l)\lambda_l^{\mathcal{N}}\mathcal{B}(\lambda_k)
-\lambda_k^{\mathcal{N}}\mathcal{B}(\lambda_l)}{(\lambda_k-\lambda_l)\mathcal{B}'(\lambda_l)}
 = \lambda_k^{\mathcal{N}} \delta_{kl},\\
&\{w_k,w_l\}_{\text{s}} =
\lim_{\substack{u\to \lambda_k \\ v\to \lambda_l}}
\left(-\frac{\mathcal{A}'(u)[f(v,u) u^{\mathcal{N}}\mathcal{B}(v) -v^{\mathcal{N}}\mathcal{B}(u)]}
{\mathcal{B}'(u)(v-u)} + \right. \\
&\phantom{\{w_k,} \left. +
\frac{\mathcal{A}'(v)[f(u,v) v^{\mathcal{N}}\mathcal{B}(u) -u^{\mathcal{N}}\mathcal{B}(v)]}
{\mathcal{B}'(v)(u-v)}\right) = -\left(\frac{\mathcal{A}'(\lambda_k)}{\mathcal{B}'(\lambda_k)} -
\frac{\mathcal{A}'(\lambda_l)}{\mathcal{B}'(\lambda_l)}\right) \lambda_k^{\mathcal{N}} \delta_{kl} = 0,
\end{split}
\end{equation*}
as required.
\end{proof}

\begin{ABLemma}\label{L:ABbracketHM}
For $\mathcal{B}$ and $\mathcal{A}$ defined by \eqref{ABNLS} the following identities are true
with respect to the second Lie-Poisson bracket \eqref{PoissonBraHM}
\begin{gather*}
\{\mathcal{B}(u),\mathcal{B}(v)\}_{\textup{s}} = 0,\quad
\{\mathcal{A}(u),\mathcal{A}(v)\}_{\textup{s}} = 0, \\
\{\mathcal{A}(u),\mathcal{B}(v)\}_{\textup{s}} = -
\frac{f(u,v) v^{\mathcal{N}}\mathcal{B}(u)-u^{\mathcal{N}}\mathcal{B}(v)}{u-v},
\end{gather*}
where $f(u,v)\,{=}\,\gamma_2^2(v)/\gamma_2^2(u)$ for \eqref{w1PolyNLS}, and
$f(u,v)\,{=}\,\gamma_3^2(v)/\gamma_3^2(u)$ for \eqref{w2PolyNLS}.
\end{ABLemma}
\begin{proof}
It repeats the proof of \nameABLemma\;\ref{L:ABbracketNLS}.

Using the second Lie-Poisson bracket in the form
$$\{L_{ij}(u), L_{kl}(v)\}_\text{f} = -\frac{v^{\mathcal{N}}L_{kj}(u)-u^{\mathcal{N}}L_{kj}(v)}{u-v}\delta_{il} +
\frac{v^{\mathcal{N}}L_{il}(u)-u^{\mathcal{N}}L_{il}(v)}{u-v} \delta_{kj},$$
one obtains
\begin{align*}
&\{\gamma_{2}(u),\Gamma_3(v)\} = -\{\gamma_{3}(u),\Gamma_2(v)\} = \frac{v^{\mathcal{N}}}{u-v} \begin{vmatrix}
\gamma_2(u) & \gamma_3(u) \\ \gamma_2(v) & \gamma_3(v) \end{vmatrix},\\
&\{\Gamma_{2}(u),\gamma_3(v)\} = -\{\Gamma_{3}(u),\gamma_2(v)\} = -\frac{u^{\mathcal{N}}}{u-v} \begin{vmatrix}
\gamma_2(u) & \gamma_3(u) \\ \gamma_2(v) & \gamma_3(v) \end{vmatrix},\\
&\{\Gamma_{2}(u),\Gamma_3(v)\} = \frac{1}{u-v}\bigg(- u^{\mathcal{N}}\begin{vmatrix}
\gamma_2(u) & \gamma_3(u) \\ \Gamma_2(v) & \Gamma_3(v) \end{vmatrix}
+ v^{\mathcal{N}}\begin{vmatrix} \gamma_2(v) & \gamma_3(v) \\ \Gamma_2(u) & \Gamma_3(u) \end{vmatrix} \bigg),
\end{align*}
and then
\begin{gather*}
\begin{split}
&\{\gamma_{2}(u),\mathcal{B}(v)\} = -\frac{v^{\mathcal{N}}\gamma_{2}(v)}{u-v} \begin{vmatrix}
\gamma_2(u) & \gamma_3(u) \\ \gamma_2(v) & \gamma_3(v) \end{vmatrix},\\
&\{\Gamma_{2}(u),\mathcal{B}(v)\} = \frac{1}{u-v} \bigg(-u^{\mathcal{N}}\gamma_{2}(u) \mathcal{B}(v)+
v^{\mathcal{N}} \gamma_{2}(v) \begin{vmatrix}
\gamma_2(v) & \gamma_3(v) \\ \Gamma_2(u) & \Gamma_3(u) \end{vmatrix} \bigg).
\end{split}\\
\Gamma_{2}(u)\{\gamma_{2}(u),\mathcal{B}(v)\} - \gamma_{2}(u)\{\Gamma_{2}(u),\mathcal{B}(v)\}
= \frac{1}{u-v} \big[\gamma^2_{2}(u) u^{\mathcal{N}}\mathcal{B}(v)-
\gamma^2_{2}(v) v^{\mathcal{N}}\mathcal{B}(u)\big].
\end{gather*}
From Leibniz's rule for a Poisson bracket with $\mathcal{A}(u)\,{=}\,{-}\Gamma_{2}(u)/\gamma_{2}(u)$
one gets
\begin{gather*}
\{\mathcal{A}(u),\mathcal{B}(v)\} = -
\frac{1}{u-v}\Big(\frac{\gamma_2^2(v)}{\gamma^2_2(u)}v^{\mathcal{N}}\mathcal{B}(u)
-u^{\mathcal{N}}\mathcal{B}(v) \Big).
\end{gather*}
Other identities from the lemma statement are computed in the similar way.
\end{proof}

(iii) The Liouville 1-form on $\mathcal{O}_{\text{s}}$ is implied by (\ref{PoissonBraHMConj}):
\begin{equation*}
\Omega_{\text{s}}=-\sum\limits_{k} \lambda^{-\mathcal{N}} w_{k}\,d\lambda_{k}.
\end{equation*}
Reduction to a Liouville torus is realized by fixing values of the
Hamiltonians $h_{\mathcal{N}}$, $h_{\mathcal{N}+1}$, \ldots, $h_{2\mathcal{N}-1}$, $f_{\mathcal{N}}$, $f_{\mathcal{N}+1}$,
\ldots, $f_{3\mathcal{N}-1}$. On the torus every $w_k$
is an algebraic function of the conjugate variable $\lambda_k$ due to (\ref{HyperCurveHM}).
After this reduction the form $\Omega_{\text{s}}$ becomes a sum of
meromorphic differentials on the Riemann surface $P(w,\lambda)=0$.

This completes the proof of \nameSoV\;\ref{T:SoVHM}. \qed
\end{proof}

Further, we construct integrable systems on the orbits $\mathcal{O}_{\text{f}}$ and $\mathcal{O}_{\text{s}}$:
a coupled 3-component nonlinear Schr\"{o}dinger equation and an isotropic SU(3) Landau-Lifshitz equation.

\section{Integrable systems on the orbits $\mathcal{O}_{\text{f}}$ and $\mathcal{O}_{\text{s}}$}

\subsection{A coupled 3-component nonlinear Schr\"{o}dinger equation}
This equation as an integrable system arises on the orbit $\mathcal{O}_{\text{f}}$
from the Hamiltonian flows generated by $h_{\mathcal{N}-2}$, $h_{\mathcal{N}-3}$.
In general, every Hamiltonian with respect to the first bracket \eqref{PoissonBraNLS}
gives rise to a nontrivial flow on $\mathcal{\mathcal{M}}_0$:
\begin{equation}\label{FlowEqNLS}
  \frac{\partial L_a^{(m)}}{\partial \tau}=\{L_a^{(m)},\mathcal{H}\}_{\text{f}},
\end{equation}
where $\mathcal{H}$ runs over the set
$\{h_0$, $h_1$, \ldots, $h_{\mathcal{N}-1}$, $f_0$, $f_1$, \ldots, $f_{2\mathcal{N}-1}\}$.
We write the flows generated by
$h_{\mathcal{N}-2}$, $h_{\mathcal{N}-3}$ in the Lax form
\begin{subequations}\label{FlowNLSEq}
\begin{gather}\label{FlowNLSEq_x}
  \frac{\partial \tens{L}(\lambda)}{\partial x}
  =[\tens{L}(\lambda),\nabla_{-1} h_{\mathcal{N}-2}]=[\nabla_{\mathcal{N}-1} h_{2\mathcal{N}-2},\tens{L}(\lambda)],\\
  \label{FlowNLSEq_t}
  \frac{\partial \tens{L}(\lambda)}{\partial t}
  =[\tens{L}(\lambda),\nabla_{-1} h_{\mathcal{N}-3}]=[\nabla_{\mathcal{N}-1} h_{2\mathcal{N}-3},\tens{L}(\lambda)],
\end{gather}
\end{subequations}
where
$\nabla_{k}$ denotes the matrix gradient with respect to the bilinear form $\langle\cdot,\cdot\rangle_{k}$:
\begin{equation*}
\nabla_{k} \mathcal{H} = \sum_{m=0}^{\mathcal{N}-1} \sum_{a=1}^{\dim \mathfrak{g}}
\frac{\partial \mathcal{H}}{\partial L_a^{(m)}} \tens{Z}_a^{k-m},
\quad \text{where} \quad L_a^{(m)} = \langle\tens{L}(\lambda),\tens{Z}_a^{k-m}\rangle_{k}.
\end{equation*}
The matrix gradients $\nabla_{\mathcal{N}-1} h_{2\mathcal{N}-2}$, $\nabla_{\mathcal{N}-1} h_{2\mathcal{N}-3}$
are used instead of $\nabla_{-1} h_{\mathcal{N}-2}$, $\nabla_{-1} h_{\mathcal{N}-3}$ due to their
simplicity:
\begin{gather*}
  \nabla_{\mathcal{N}-1} h_{2\mathcal{N}-2} =
  \lambda \tens{L}^{(\mathcal{N})} + \tens{L}^{(\mathcal{N}-1)},\quad
  \nabla_{\mathcal{N}-1} h_{2\mathcal{N}-3} =
  \lambda^2 \tens{L}^{(\mathcal{N})} + \lambda\tens{L}^{(\mathcal{N}-1)}
  + \tens{L}^{(\mathcal{N}-2)},
\end{gather*}
where $\tens{L}^{(m)}$ denotes the matrix coefficient of $\tens{L}$ of power $m$.

The chosen Hamiltonian flows commute, that can be expressed through the compatibility condition
which is the zero curvature equation
\begin{gather*}
\frac{\partial \nabla_{-1} h_{\mathcal{N}-2}}{\partial t} -
\frac{\partial \nabla_{-1} h_{\mathcal{N}-3}}{\partial x} +
[\nabla_{-1} h_{\mathcal{N}-3},\nabla_{-1} h_{\mathcal{N}-2}] = 0.
\end{gather*}

Now we show how to obtain a coupled 3-component nonlinear Schr\"{o}dinger equation
form (\ref{FlowNLSEq}). The coefficients of power $\mathcal{N}$ in $\lambda$
displays that $\tens{L}^{(\mathcal{N})}$ is constant in $x$ and $t$,
we assign
$\tens{L}^{(\mathcal{N})}\,{=}\,\diag \big(\alpha_1^{(\mathcal{N})}$,
$\alpha_2^{(\mathcal{N})}\,{-}\,\alpha_1^{(\mathcal{N})}$, ${-}\alpha_2^{(\mathcal{N})}\big)$.
The coefficient of power $\mathcal{N}\,{-}\,1$ from (\ref{FlowNLSEq_x}) gives expressions for
the dynamic variables $\{L_a^{(\mathcal{N}-2)}\}$ in terms of $\{L_a^{(\mathcal{N}-1)},\,\partial_x L_a^{(\mathcal{N}-1)}\}$:
\begin{align*}
&\beta_1^{(\mathcal{N}-2)} = \frac{\partial_x \beta_1^{(\mathcal{N}-1)}}{2\alpha_1^{(\mathcal{N})}- \alpha_2^{(\mathcal{N})}}&
&\gamma_1^{(\mathcal{N}-2)} = - \frac{\partial_x \gamma_1^{(\mathcal{N}-1)}}{2\alpha_1^{(\mathcal{N})}- \alpha_2^{(\mathcal{N})}}&\\
&\beta_2^{(\mathcal{N}-2)} = \frac{\partial_x \beta_2^{(\mathcal{N}-1)}}{2\alpha_2^{(\mathcal{N})} - \alpha_1^{(\mathcal{N})}}&
&\gamma_2^{(\mathcal{N}-2)} = - \frac{\partial_x \gamma_2^{(\mathcal{N}-1)}}{2\alpha_2^{(\mathcal{N})}- \alpha_1^{(\mathcal{N})}}&\\
&\beta_3^{(\mathcal{N}-2)} = \frac{\partial_x \beta_3^{(\mathcal{N}-1)}}{\alpha_1^{(\mathcal{N})}+ \alpha_2^{(\mathcal{N})}}&
&\gamma_3^{(\mathcal{N}-2)} = - \frac{\partial_x \gamma_3^{(\mathcal{N}-1)}}{\alpha_1^{(\mathcal{N})} + \alpha_1^{(\mathcal{N})}}.&
\end{align*}
The rest of equations show that $\alpha_1^{(\mathcal{N}-1)}$, $\alpha_2^{(\mathcal{N}-1)}$ are constant,
we compute them from the orbit equations
$h_{2\mathcal{N}-1}\,{=}\,c_{2\mathcal{N}-1}$, $f_{3\mathcal{N}-1}\,{=}\,d_{3\mathcal{N}-1}$.
The coefficient of power $\mathcal{N}\,{-}\,2$ from (\ref{FlowNLSEq_x}) allows to
express $\{L_a^{(\mathcal{N}-3)}\}$
in terms of $\{L_a^{(\mathcal{N}-1)}$, $\partial_x L_a^{(\mathcal{N}-1)}$, $\partial_{xx} L_a^{(\mathcal{N}-1)}\}$,
in particular:
\begin{equation}\label{bEq}
\begin{split}
&\big(2\alpha_1^{(\mathcal{N})}- \alpha_2^{(\mathcal{N})}\big) \beta_1^{(\mathcal{N}-3)} = \frac{\partial^2_{xx} \beta_1^{(\mathcal{N}-1)}}{2\alpha_1^{(\mathcal{N})}- \alpha_2^{(\mathcal{N})}}
-\frac{\big(2\alpha_1^{(\mathcal{N}-1)}-\alpha_2^{(\mathcal{N}-1)}\big) \partial_{x}\beta_1^{(\mathcal{N}-1)}} {2\alpha_1^{(\mathcal{N})}-\alpha_2^{(\mathcal{N})}}
+ \\
&\phantom{\quad\quad\quad} +\frac{\beta_3^{(\mathcal{N}-1)}\partial_{x}\gamma_2^{(\mathcal{N}-1)}}
{2\alpha_2^{(\mathcal{N})}-\alpha_1^{(\mathcal{N})}} + \frac{\gamma_2^{(\mathcal{N}-1)}\partial_{x}\beta_3^{(\mathcal{N}-1)}}
{\alpha_1^{(\mathcal{N})}+\alpha_2^{(\mathcal{N})}}
+ \big(2\alpha_1^{(\mathcal{N}-2)}- \alpha_2^{(\mathcal{N}-2)}\big) \beta_1^{(\mathcal{N}-1)},\\
&\big(2\alpha_2^{(\mathcal{N})}- \alpha_1^{(\mathcal{N})}\big) \beta_2^{(\mathcal{N}-3)} = \frac{\partial^2_{xx} \beta_2^{(\mathcal{N}-1)}}{2\alpha_2^{(\mathcal{N})} - \alpha_1^{(\mathcal{N})}}
-\frac{\big(2\alpha_2^{(\mathcal{N}-1)}-\alpha_1^{(\mathcal{N}-1)}\big) \partial_{x}\beta_2^{(\mathcal{N}-1)}} {2\alpha_2^{(\mathcal{N})}-\alpha_1^{(\mathcal{N})}}
-\\
&\phantom{\quad\quad\quad} - \frac{\beta_3^{(\mathcal{N}-1)}\partial_{x}\gamma_1^{(\mathcal{N}-1)}}
{2\alpha_1^{(\mathcal{N})}-\alpha_2^{(\mathcal{N})}} - \frac{\gamma_1^{(\mathcal{N}-1)}\partial_{x}\beta_3^{(\mathcal{N}-1)}}
{\alpha_1^{(\mathcal{N})}+\alpha_2^{(\mathcal{N})}}
+ \big(2\alpha_2^{(\mathcal{N}-2)}- \alpha_1^{(\mathcal{N}-2)}\big) \beta_2^{(\mathcal{N}-1)},\\
&\big(\alpha_1^{(\mathcal{N})}+\alpha_2^{(\mathcal{N})}\big) \beta_3^{(\mathcal{N}-3)} = \frac{\partial^2_{xx} \beta_3^{(\mathcal{N}-1)}}{\alpha_1^{(\mathcal{N})}+\alpha_2^{(\mathcal{N})}}
-\frac{\big(\alpha_1^{(\mathcal{N}-1)}+\alpha_2^{(\mathcal{N}-1)}\big) \partial_{x}\beta_3^{(\mathcal{N}-1)}} {\alpha_1^{(\mathcal{N})}+\alpha_2^{(\mathcal{N})}}
-\\
&\phantom{\quad\quad\quad} - \frac{\beta_1^{(\mathcal{N}-1)}\partial_{x}\beta_2^{(\mathcal{N}-1)}}
{2\alpha_2^{(\mathcal{N})}-\alpha_1^{(\mathcal{N})}}+
\frac{\beta_2^{(\mathcal{N}-1)}\partial_{x}\beta_1^{(\mathcal{N}-1)}}
{2\alpha_1^{(\mathcal{N})}-\alpha_2^{(\mathcal{N})}}
+ \big(\alpha_1^{(\mathcal{N}-2)}+\alpha_2^{(\mathcal{N}-2)}\big) \beta_3^{(\mathcal{N}-1)}.
\end{split}
\end{equation}
The equations for $\alpha_1^{(\mathcal{N}-2)}$ and $\alpha_2^{(\mathcal{N}-2)}$
are  easily integrated and give
\begin{align*}
&\alpha_1^{(\mathcal{N}-2)} = -\frac{\beta_1^{(\mathcal{N}-1)} \gamma_1^{(\mathcal{N}-1)}}
{2\alpha_1^{(\mathcal{N})}-\alpha_2^{(\mathcal{N})}}
-\frac{\beta_3^{(\mathcal{N}-1)} \gamma_3^{(\mathcal{N}-1)}}
{\alpha_1^{(\mathcal{N})}+\alpha_2^{(\mathcal{N})}} + C_1,\\
&\alpha_2^{(\mathcal{N}-2)} = -\frac{\beta_2^{(\mathcal{N}-1)} \gamma_2^{(\mathcal{N}-1)}}
{2\alpha_2^{(\mathcal{N})}-\alpha_1^{(\mathcal{N})}}
-\frac{\beta_3^{(\mathcal{N}-1)} \gamma_3^{(\mathcal{N}-1)}}
{\alpha_1^{(\mathcal{N})}+\alpha_2^{(\mathcal{N})}} + C_2.
\end{align*}
The constants $C_1$, $C_2$ are computed from the orbit equations
$h_{2\mathcal{N}-2}\,{=}\,c_{2\mathcal{N}-2}$, $f_{3\mathcal{N}-2}\,{=}\,d_{3\mathcal{N}-2}$.

\begin{remark}
From (\ref{FlowNLSEq_x}) one obtains expressions for all dynamic variables in
terms of $\{L_a^{(\mathcal{N}-1)}\}$ and their derivatives with respect to $x$.
\end{remark}
Next, we write the coefficient of power $\mathcal{N}\,{-}\,1$ from (\ref{FlowNLSEq_t}):
\begin{gather*}
\frac{\partial \beta_1^{(\mathcal{N}-1)}}{\partial t} = \big(2\alpha_1^{(\mathcal{N})}- \alpha_2^{(\mathcal{N})}\big)\beta_1^{(\mathcal{N}-3)}\qquad
\frac{\partial \gamma_1^{(\mathcal{N}-1)}}{\partial t} = -\big(2\alpha_1^{(\mathcal{N})}- \alpha_2^{(\mathcal{N})}\big)\gamma_1^{(\mathcal{N}-3)}\\
\frac{\partial \beta_2^{(\mathcal{N}-1)}}{\partial t} = \big(2\alpha_2^{(\mathcal{N})}- \alpha_1^{(\mathcal{N})}\big)\beta_2^{(\mathcal{N}-3)}\qquad
\frac{\partial \gamma_2^{(\mathcal{N}-1)}}{\partial t} = -\big(2\alpha_2^{(\mathcal{N})}- \alpha_1^{(\mathcal{N})}\big)\gamma_2^{(\mathcal{N}-3)}\\
\frac{\partial \beta_3^{(\mathcal{N}-1)}}{\partial t} = \big(\alpha_1^{(\mathcal{N})}+ \alpha_2^{(\mathcal{N})}\big)\beta_3^{(\mathcal{N}-3)}\qquad
\frac{\partial \gamma_3^{(\mathcal{N}-1)}}{\partial t} = -\big(\alpha_1^{(\mathcal{N})}+ \alpha_2^{(\mathcal{N})}\big)\gamma_3^{(\mathcal{N}-3)}
\end{gather*}
and substitute the expressions from \eqref{bEq} instead of $\{L_a^{(\mathcal{N}-3)}\}$.

Assigning $\alpha_{1,2}^{(m)}\,{=}\, \rmi a_{1,2}^{(m)}$,
$\gamma_{1,2,3}^{(m)}\,{=}\,{-}\big(\beta_{1,2,3}^{(m)}\big)^{*}$ we restrict the system to
an $\mathfrak{su}(3)$ loop algebra,
and put $c_{2\mathcal{N}-1}= d_{3\mathcal{N}-1}\,{=}\,c_{2\mathcal{N}-2}\,{=}\,d_{3\mathcal{N}-2}\,{=}\,0$ implying $a_1^{(\mathcal{N}-1)}\,{=}\,a_2^{(\mathcal{N}-1)}\,{=}\,0$
and  $C_1\,{=}\,C_2\,{=}\,0$.
The final equations for
$\beta_{1,2,3}^{(\mathcal{N}-1)}$ (the superscripts are omitted, namely:
$\beta_{1,2,3}^{(\mathcal{N}-1)}\,{=}\,\beta_{1,2,3}$, $a_{1,2}^{(\mathcal{N})}\,{=}\,a_{1,2}$)
\begin{equation}\label{3cNLS}
\begin{split}
&\rmi\frac{\partial \beta_1}{\partial t} =
\frac{\partial^2_{xx} \beta_1}{2a_1- a_2}
-\frac{\beta_3 \partial_{x} \beta_2^{\ast}}{\big(2a_2-a_1\big)}
-\frac{\beta_2^{\ast} \partial_{x} \beta_3}{\big(a_1+a_2\big)} + \\
&\phantom{\rmi\frac{\partial \beta_1}{\partial t} =}
+\bigg(\frac{2|\beta_1|^2}{2a_1-a_2}
+\frac{|\beta_3|^2}{a_1+a_2}
-\frac{|\beta_2|^2}{2a_2-a_1} \bigg) \beta_1,\\
&\rmi\frac{\partial \beta_2}{\partial t} =
\frac{\partial^2_{xx} \beta_2}{2a_2 - a_1}
+\frac{\beta_3\partial_{x}\beta_1^{\ast}}{\big(2a_1-a_2\big)}
+\frac{\beta_1^{\ast} \partial_{x}\beta_3}{\big(a_1+a_2\big)} + \\
&\phantom{\rmi\frac{\partial \beta_2}{\partial t} =}
+\bigg(\frac{2|\beta_2|^2}{2a_2-a_1}
+\frac{|\beta_3|^2}{a_1+a_2}-
\frac{|\beta_1|^2}{2a_1-a_2}\bigg) \beta_2,\\
&\rmi\frac{\partial \beta_3}{\partial t} =
\frac{\partial^2_{xx} \beta_3}{a_1+a_2}
-\frac{\beta_1\partial_{x}\beta_2}{\big(2a_2-a_1\big)}
 +\frac{\beta_2\partial_{x}\beta_1}{\big(2a_1-a_2\big)} + \\
&\phantom{\rmi\frac{\partial \beta_3}{\partial t} =}
+ \bigg(\frac{|\beta_1|^2}{2a_1-a_2}
+\frac{|\beta_2|^2}{2a_2-a_1}
+\frac{2|\beta_3|^2}{a_1+a_2}\bigg) \beta_3
\end{split}
\end{equation}
are the same as presented in \cite{FordyKulish}, called there
the `3-wave hierarchy generalization
of the nonlinear Schr\"{o}\-din\-ger equation'.
Here we call them a \emph{coupled 3-component nonlinear Schr\"{o}\-din\-ger equation}.

\subsection{An isotropic SU(3) Landau-Lifshitz equation}
This equation arises on the orbit $\mathcal{O}_{\text{s}}$
from the Hamiltonian flows generated by $h_{\mathcal{N}}$, $h_{\mathcal{N}+1}$.
With respect to the second bracket \eqref{PoissonBraHM} every Hamiltonian $\mathcal{H}$ from the set
$\{h_{\mathcal{N}}$, $h_{\mathcal{N}+1}$, \ldots, $h_{2\mathcal{N}-1}$, $f_{\mathcal{N}}$,
$f_{\mathcal{N}+1}$, \ldots, $f_{3\mathcal{N}-1}\}$
gives rise to a nontrivial flow on $\mathcal{\mathcal{M}}_0$:
\begin{equation}\label{FlowEqHM}
  \frac{\partial L_a^{(m)}}{\partial \tau}=\{L_a^{(m)},\mathcal{H}\}_{\text{s}}.
\end{equation}
We write the flows of $h_{\mathcal{N}}$, $h_{\mathcal{N}+1}$ in the Lax form:
\begin{subequations}\label{FlowHMEq}
\begin{gather}
  \frac{\partial \tens{L}(\lambda)}{\partial x}=
  [\nabla_{\mathcal{N}-1} h_{\mathcal{N}},\tens{L}(\lambda)]=
  [\tens{L}(\lambda),\nabla_{-1} h_0], \\
   \frac{\partial \tens{L}(\lambda)}{\partial t}=
   [\nabla_{\mathcal{N}-1} h_{\mathcal{N}+1},\tens{L}(\lambda)]=
   [\tens{L}(\lambda),\nabla_{-1} h_1],
\end{gather}
\end{subequations}
where for the sake of simplicity we use expressions with the matrix gradients
\begin{gather*}
  \nabla_{-1} h_0 = \lambda^{-1}\,\tens{L}^{(0)}, \qquad
  \nabla_{-1} h_1 = \lambda^{-1}\,\tens{L}^{(1)} + \lambda^{-2}\,\tens{L}^{(0)}.
\end{gather*}

This system is also constructed in the loop algebra $\mathfrak{su}(3)$, and we change the basis $\{\tens{Z}_a\}$ into
the Gell-Mann basis $\{\tens{X}_a\,{:}\,a=1,\,\dots,\,8\}$ (see \cite{Macfarlane}):
\begin{gather*}
\Tr \tens{X}_a \tens{X}_b = -\tfrac{1}{2}\,\delta_{ab},\quad
[\tens{X}_a,\tens{X}_b] = \textsl{f}_{abc} \tens{X}_c,\quad
\tens{X}_a \tens{X}_b + \tens{X}_b \tens{X}_a = -\tfrac{1}{3}\,\delta_{ab} \Ibb - \tfrac{3}{2}\, \textsl{d}_{abc} X_c,\\
\textsl{f}_{abc} = -2 \Tr \tens{X}_c [\tens{X}_a,\tens{X}_b],\, \quad
\textsl{d}_{abc} =  \tfrac{4}{3}\,\Tr \tens{X}_c \big(\tens{X}_a \tens{X}_b + \tens{X}_b \tens{X}_a\big).
\end{gather*}
As above for dynamic variables we use the coordinates corresponding to the basis elements
with respect to the bilinear form, that is $\{\mu_a^{(m)}\,{=}\, \langle\tens{L}(\lambda),
\tens{X}_a^{\mathcal{N}-1-m}\rangle_{\mathcal{N}-1}\}$ serve as dynamic variables for the system
on the orbit $\mathcal{O}_{\text{s}}$. The matrix coefficient of $\tens{L}$ of power $m$ has the form
\begin{gather*}
\small \tens{L}^{(m)} = \rmi \begin{pmatrix}
\mu_3^{(m)} + \frac{1}{\sqrt{3}}\,\mu_8^{(m)} & \mu_1^{(m)} - \rmi \mu_2^{(m)} & \mu_4^{(m)} - \rmi \mu_5^{(m)}\\
\mu_1^{(m)} + \rmi \mu_2^{(m)} & -\mu_3^{(m)} + \frac{1}{\sqrt{3}}\,\mu_8^{(m)} & \mu_6^{(m)} - \rmi \mu_7^{(m)}\\
\mu_4^{(m)} + \rmi \mu_5^{(m)} & \mu_6^{(m)} + \rmi \mu_7^{(m)} & - \frac{2}{\sqrt{3}}\,\mu_8^{(m)}
\end{pmatrix},
\end{gather*}
which allows to get relations between $\{\mu_a^{(m)}\}$ and $\{\alpha_{1,2}^{(m)}$, $\beta_{1,2,3}^{(m)}$, $\gamma_{1,2,3}^{(m)}\}$. The Poisson structure in terms of the new dynamic variables is given by
\begin{equation*}
\{\mu_a^{(m)},\mu_b^{(n)}\}_\text{s} = -\textsl{f}_{abc} \mu_c^{(m+n+1-\mathcal{N})}.
\end{equation*}
The orbit $\mathcal{O}_\text{s}$ is defined by the equations (in matrix and vector notations)
\begin{gather*}
  \begin{array}{llll}
  \frac{1}{2}\Tr \big(\tens{L}^{(0)}\big)^2 = - \mu_a^{(0)}\mu_a^{(0)} =c_{0}
  &\ \ && \frac{1}{3}\Tr \big(\tens{L}^{(0)}\big)^3 =  -\textsl{d}_{abc} \mu_a^{(0)} \mu_b^{(0)} \mu_c^{(0)} = d_0\\
  \Tr \tens{L}^{(0)} \tens{L}^{(1)} = -2 \mu_a^{(0)} \mu_a^{(1)} = c_{1}
  &\ \ && \Tr \tens{L}^{(0)} \tens{L}^{(0)} \tens{L}^{(1)} = -3 \textsl{d}_{abc} \mu_a^{(0)} \mu_b^{(0)} \mu_c^{(1)} = d_1 \\
  .\ .\ .\ .\ .\ .\ .\ .\ .\ .\ .\ .\ .\ .\ .\ .\ .\ .
  &\ \ && .\ .\ .\ .\ .\ .\ .\ .\ .\ .\ .\ .\ .\ .\ .\ .\ .\ .\ .\ .\ .\ .\ .\ .\  \\
  \frac{1}{2}\, \Tr \sum\limits_{m+n=\mathcal{N}} \tens{L}^{(m)} \tens{L}^{(n)} = &\ \ &&
  \frac{1}{3}\, \Tr \sum\limits_{m+n+k=\mathcal{N}} \tens{L}^{(m)} \tens{L}^{(n)} \tens{L}^{(k)}  = \\
  \phantom{=} =-\sum\limits_{m+n=\mathcal{N}} \mu_a^{(m)} \mu_a^{(n)} = c_{\mathcal{N}}, &\ \ &&
  \phantom{=} = -\sum\limits_{m+n+k=\mathcal{N}}  \textsl{d}_{abc} \mu_a^{(m)} \mu_b^{(n)} \mu_c^{(k)} = d_{\mathcal{N}},
  \end{array}
 \end{gather*}
where indices appearing twice imply summation.
The Hamiltonian equations obtained from (\ref{FlowHMEq}) have the following vector and matrix forms:
\begin{subequations}
\renewcommand{\theequation}{\theparentequation\alph{equation}}
\begin{gather} \label{FlowHMEqx}
  \frac{\partial \mu_a^{(m)}}{\partial x} \,{=}\,{-} 2\textsl{f}_{abc} \mu_b^{(m+1)}\mu_c^{(0)}, \
  \frac{\partial \mu_a^{(m)}}{\partial t} \,{=}\,{-} 2\textsl{f}_{abc}
  \big(\mu_b^{(m+1)}\mu_c^{(1)} + \mu_b^{(m+2)}\mu_c^{(0)}\big) \\ \label{FlowHMEqt}
  \frac{\partial \tens{L}^{(m)}}{\partial x} =  [\tens{L}^{(m+1)},\tens{L}^{(0)}],
  \qquad
  \frac{\partial \tens{L}^{(m)}}{\partial t} = [\tens{L}^{(m+1)},\tens{L}^{(1)}]
  + [\tens{L}^{(m+2)},\tens{L}^{(0)}].
\end{gather}
\end{subequations}
We denote the variables $\{\mu_a^{(0)}\}$ by $\{\mu_a\}$, and $\tens{L}^{(0)}$ by $\tens{M}$, and introduce the variables
$\{T_a\,{=}\,\textsl{d}_{abc} \mu_b \mu_c\}$ or in the matrix form $\tens{T}\,{=}\,\frac{2}{3}\,\tens{M}^2\,{-}\,\frac{4}{9}\,c_0 \Ibb$.
The stationary equation at $m\,{=}\,0$ looks like
$$\frac{\partial \tens{M}}{\partial x} \,{=}\,\ad_{\tens{L}^{(1)}} \tens{M},\qquad  \tens{M},\, \tens{L}^{(1)} \in \mathfrak{g}^\ast,$$
where $\ad$ is the adjoint action in the Lie algebra $\mathfrak{g}$. This equation
can be easily solved on the orbit $\mathcal{O}_\text{s}$, where  $\tens{M}^3\,{=}\,c_0 \tens{M} \,{+}\, d_0\Ibb$
or in the vector form $\textsl{d}_{abc} \textsl{d}_{cpq} \mu_b\mu_p\mu_q\,{=}\,\frac{4}{27}\,c_0\mu_a$, namely:
\begin{gather*}
\mu^{(1)}_a = \tfrac{2}{4c_0^3 - 27 d_0^2}\, \textsl{f}_{abc} \Big(
 \tfrac{27}{4}\, c_0 T_b T_{c,x} - \tfrac{27}{2}\, d_0 T_b \mu_{c,x} + c_0^2 \mu_b \mu_{c,x}\Big)
+ \\ + \tfrac{2c_1 c_0^2 - 9 d_1 d_0}{4c_0^3 - 27 d_0^2}\, \mu_a +
\tfrac{9}{2}\tfrac{2 d_1 c_0 - 3 c_1 d_0}{4c_0^3 - 27 d_0^2}\, T_a  \quad
\text{or}\\
\tens{L}^{(1)} =  \tfrac{-1}{4c_0^3 - 27 d_0^2}\Big( \tfrac{27}{4}\,c_0 [\tens{T},\tens{T}_x] -
\tfrac{27}{2}\,d_0 [\tens{T},\tens{M}_x] + c_0^2 [\tens{M},\tens{M}_x]\Big) + \\ +
\tfrac{2c_1 c_0^2 - 9 d_1 d_0}{4 c_0^3-27 d_0^2}\, \tens{M}
+ \tfrac{9}{2}\tfrac{ 2 d_1 c_0 - 3 c_1 d_0 }{4 c_0^3-27 d_0^2} \,\tens{T} .
\end{gather*}
Commutativity of the chosen Hamiltonian flows implies
the compatibility condition in the zero curvature form
\begin{gather*}
\frac{\partial \nabla_{\mathcal{N}-1} h_{\mathcal{N}}}{\partial t} -
\frac{\partial \nabla_{\mathcal{N}-1} h_{\mathcal{N}+1}}{\partial x} +
[\nabla_{\mathcal{N}-1} h_{\mathcal{N}},\nabla_{\mathcal{N}-1} h_{\mathcal{N}+1}] = 0,
\end{gather*}
that gives in particular
\begin{equation*}
    \frac{\partial \mu_a }{\partial t}=
    \frac{\partial \mu_a^{(1)} }{\partial x}\qquad \text{or} \qquad
    \frac{\partial \tens{M} }{\partial t}=
    \frac{\partial \tens{L}^{(1)} }{\partial x}.
\end{equation*}
Then we get the equation
\begin{subequations}\label{HMEq}
\begin{gather}
    \frac{\partial \mu_a }{\partial t}=\tfrac{2}{4c_0^3 - 27 d_0^2}\,
\textsl{f}_{abc} \Big( \tfrac{27}{4}\, c_0 T_b T_{c,xx} - \tfrac{27}{2}\, d_0 [T_b \mu_{c,xx} + \mu_b T_{c,xx}]
+ c_0^2 \mu_b \mu_{c,xx}\Big) + \notag \\ + \tfrac{2c_1 c_0^2 - 9 d_1 d_0}{4 c_0^3-27 d_0^2}\, \mu_{a,x} +
\tfrac{9}{2}\tfrac{ 2 d_1 c_0 - 3 c_1 d_0 }{4 c_0^3-27 d_0^2}  T_{a,x} \quad \text{or} \\
    \frac{\partial \tens{M}}{\partial t} =
    \tfrac{-1}{4c_0^3 - 27 d_0^2}\Big( \tfrac{27}{4}\,c_0 [\tens{T},\tens{T}_{xx}] -
\tfrac{27}{2}\,d_0 ([\tens{T}_x,\tens{M}_{x}] + [\tens{T},\tens{M}_{xx}])
+ c_0^2 [\tens{M},\tens{M}_{xx}]\Big) + \notag
\\ \tfrac{2c_1 c_0^2 - 9 d_1 d_0}{4 c_0^3-27 d_0^2}\, \tens{M}_x +
\tfrac{9}{2}\tfrac{ 2 d_1 c_0 - 3 c_1 d_0 }{4 c_0^3-27 d_0^2} \,\tens{T}_x.
\end{gather}
\end{subequations}
The obtained equation is similar to the Landau-Lifshitz equation for an isotropic SU(2) magnet, and so we call it
a \emph{generalized Landau-Lifshitz equation for an
isotropic SU(3) magnet}, for more details see \cite{BernHolod09}.

\section{Conclusion and Discussion}
We briefly summarize the proposed separation of variables procedure.
Recall that we deal with an integrable system constructed on a coadjoint orbit
of a loop Lie algebra, and we use the Cartan-Weyl basis.
The key point of the proposed procedure is restriction to an orbit located in the dual space
to the loop algebra. We realize this restriction by eliminating a subset of dynamic
variables corresponding to nilpotent commuting basis elements.
On the other hand, we parameterize the orbit by a sufficient number of points of the spectral curve
$\det \bigl(\tens{L}(\lambda)\,{-}\,w\Ibb\bigr)\,{=}\,0$, where
$\tens{L}$ is the Lax matrix of the system.
Thus, we obtain two representations for every point of the orbit: in the dynamic
and the spectral variables. It is possible to introduce these variables
so that the map between them is biunique.
The spectral variables are proven to be variables of separation.

In our opinion, the orbit approach allows to
`elucidate the geometric and algebraic meaning of the construction', that was
declared as an unsolved problem in \cite{Sklyanin92}. Moreover,
the procedure can be easily extended
to generic orbits of $\mathfrak{sl}(n)$ loop algebras, the only problem is cumbersome computations.
Another open question is an extension of the orbit approach to degenerate orbits.
Though the latter have a simpler geometry than generic orbits, it is difficult to
define them by explicit equations, that causes problems with applying the proposed procedure.


\label{lastpage}

\end{document}